\documentclass[longauth,letter]{aa}     

\usepackage{graphicx}
\usepackage{txfonts}
\usepackage{lipsum}

\usepackage{paracol}
\usepackage{amsmath}
\usepackage{color}

\usepackage{subcaption}         
\usepackage{lscape}             
\usepackage{placeins}           

\usepackage[colorlinks=true, linkcolor=blue, citecolor=blue, urlcolor=blue]{hyperref}

\begin{document}
   \title{EP251023a: A fast X-ray transient featuring a magnetar-powered optical internal plateau followed by a steep decay}

%
%
%

   \author{Shuai-Qing Jiang\inst{1,2}\email{sqjiang@bao.ac.cn}       
        \and Dong Xu\inst{1,3}\corrauth{dxu@nao.cas.cn} 
        \and Wei-Hua Lei\inst{4}\corrauth{leiwh@hust.edu.cn}
        \and Jie An\inst{1,2}\email{anjie@bao.ac.cn}
        \and Yuan-Chuan Zou\inst{4}\email{zouyc@hust.edu.cn}
        \and Zi-Pei Zhu\inst{1}\email{zpzhu@nao.cas.cn}
        \and Ryan~Chornock\inst{5,6}\email{chornock@berkeley.edu}
        \and Dmitry Svinkin\inst{7}\email{svinkin@mail.ioffe.ru}
        \and Wen-Xiong Li\inst{1}\email{liwx@bao.ac.cn}
        \and E. Fern\'andez-Garc\'ia\inst{8}\email{emifdez@iaa.es}
        \and Yue Wu\inst{9}\email{dz21260011@smail.nju.edu.cn}
        \and Wen-Da~Zhang\inst{1}\email{wdzhang@nao.cas.cn}
        \and Shao-Yu Fu\inst{4}\email{syfu@nao.cas.cn}
        \and Xing~Liu\inst{1,2}\email{liuxing@nao.cas.cn}
        \and Lin-Bo~He\inst{1,2}\email{helb@bao.ac.cn}
        \and Moira Andrews\inst{10,11}\email{mandrews@lco.global}
        \and A. J. Castro-Tirado\inst{8,12}\email{ajct@iaa.es}
        \and Joseph R. Farah\inst{10,11}\email{jfarah@lco.global}
        \and Dmitry Frederiks\inst{7}\email{fred@mail.ioffe.ru}
        \and M.~Gritsevich\inst{8,13,14}\email{maria@iaa.es}
        \and D. Andrew Howell\inst{10,11}\email{ahowell@lco.global}
        \and Ding-Fang Hu\inst{15}\email{dfhu@pmo.ac.cn}
        \and Alexandra L. Lysenko\inst{7}\email{Alexandra.Lysenko@mail.ioffe.ru}
        \and A. Maury\inst{16}\email{amaury2020@spaceobs.com}
        \and Curtis McCully\inst{10}\email{cmccully@lco.global}
        \and S. B. Pandey\inst{17}\email{shashiaries0@gmail.com}
        \and I. P\'erez-Garc\'ia\inst{8}\email{ipg@iaa.es}
        \and Anna Ridnaia\inst{7}\email{ridnaia@mail.ioffe.ru}
        \and Anastasia Tsvetkova\inst{7}\email{tsvetkova@mail.ioffe.ru}
        \and Mikhail Ulanov\inst{7}\email{ulanov@mail.ioffe.ru}
        \and S.-Y. Wu\inst{8}\email{wusiyu.11@outlook.com}
        \and Kathryn~Wynn\inst{10,11}\email{kwynn@lco.global}
        \and D.-R. Xiong\inst{18}\email{xiongdingrong@ynao.ac.cn}
        \and Hao-Nan Yang\inst{1}\email{hnyang@nao.cas.cn}
        \and B.-B. Zhang\inst{9,19}\email{bbzhang@nju.edu.cn}
        \and Tong Zhao\inst{1}\email{zhaotong@bao.ac.cn}
        }

   \institute{National Astronomical Observatories, Chinese Academy of Sciences, Beijing 100101, People's Republic of China
   \and School of Astronomy and Space Science, University of Chinese Academy of Sciences, Chinese Academy of Sciences, Beijing 100049, People's Republic of China
   \and Altay Astronomical Observatory, Altay, Xinjiang 836500, People's Republic of China
   \and Department of Astronomy, School of Physics, Huazhong University of Science and Technology, Wuhan, 430074, People’s Republic of China
   \and Department of Astronomy, University of California, 
Berkeley, CA 94720-3411, USA
   \and Berkeley Center for Multi-messenger Research on 
Astrophysical Transients and Outreach (Multi-RAPTOR), University of 
California, Berkeley, CA 94720-3411, USA
   \and Ioffe Institute, Politekhnicheskaya 26, St. Petersburg, 194021, Russia 
   \and Instituto de Astrof\'isica de Andaluc\'ia (IAA-CSIC), Glorieta  
de la Astronom\'ia s/n, E-18008, Granada
   \and School of Astronomy and Space Science, Nanjing University, Nanjing 210093, People's Republic of China
   \and Las Cumbres Observatory, 6740 Cortona Drive Suite 102, Goleta, CA 93117-5575, USA
   \and Department of Physics, University of California, Santa Barbara, CA 93106-9530, USA
   \and Unidad Asociada al CSIC Departamento de Ingenier\'ia de  
Sistemas y Autom\'atica, Escuela de Ingenier\'ias, Universidad de  
M\'alaga, Dr. Ortiz Ramos s\/n, E-29071, M\'alaga, Spain
   \and Department of Physics, University of Helsinki, Gustav  
H\"allstr\"omin katu 2, FI-00014 Helsinki, Finland
   \and Institute of Physics and Technology, Ural Federal University,  
Mira street 19, 620002 Ekaterinburg, Russia
   \and Purple Mountain Observatory, Chinese Academy of Sciences, Nanjing 210023, People's Republic of China 
   \and SPACE Celestial Explorations, San Pedro de Atacama, Chile
   \and Aryabhatta Research Institute of Observational Sciences  
(ARIES), Manora Peak, Nainital-263002, India
   \and Kunming Astronomical Observatory, Yunnan Astronomical  
Observatory, China Academy of Sciences, 2QCR+V2R, Z031, Guandu  
District, 650208 Kunming, Yunnan, China
   \and Key Laboratory of Modern Astronomy and Astrophysics (Nanjing  
University), Ministry of Education, Nanjing 210093, China
}


 
  \abstract
{EP251023a is an extragalactic fast X-ray transient (eFXT) detected solely by EP without a gamma-ray counterpart. The prompt emission consists of a main emission with a duration $T_{90}=292\pm19$ s, followed by a long-lasting tail emission that persists until the observation ends at $T_0+1571$ s. With the upper limit of Konus--Wind, we derived a conservative upper limit on the isotropic gamma-ray energy $E_{\gamma,\rm{iso}}$ of $5.7 \times 10^{52}$ erg for the main emission phase. A redshift of $z = 2.232\pm0.001$ is identified from strong absorption features in the Keck spectrum, which also indicate a relatively low host-galaxy HI column density. Based on the broadband spectral energy distribution, the late-time light curves show an achromatic plateau, followed by an extremely steep decay with a slope of 3.99 after a break at about 49 ks, which is consistent with a rapidly spinning millisecond magnetar engine. Under the isotropic wind scenario, we obtain the initial period $P_0<2.27$~ms and the magnetic field strength $B_p<8.33\times10^{14}$~G for the magnetar; whereas considering a jet collimation with a typical opening angle of 0.1 rad relaxes these constraints to $P_0<32.15$~ms and $B_p<1.18\times10^{16}$~G. Together with GRB\,070707, EP251023a may represent a rare class of optical magnetar-powered internal plateaus with little external-shock contamination, unlike previous examples detected primarily in X-rays. Future discoveries of similar events will help clarify the relationship between magnetar-powered internal emission observed in the optical band and that detected only in X-rays.}


   \titlerunning{EP251023a}
   \authorrunning{Shuai-Qing Jiang et al.}
   \maketitle

\section{Introduction}\label{intro}
A long-lived central engine, such as a millisecond magnetar \citep{Zhang_2001}, is often invoked to explain the shallow decay (plateau) followed by a steep drop in gamma-ray burst (GRB) afterglows. This phenomenology is well-established in the X-ray band for various GRBs (e.g., \citealt{Rowlinson_2013}) and X-ray transients like CDF-S XT2 \citep{Xue_nature}.
Beyond the X-ray band, such plateau-to-steep-decay features are also observed in optical afterglows, which are typically interpreted as internal plateaus. Nevertheless, for most events, isolating the optical internal plateau emission requires subtracting contributions from other components, which introduces a certain degree of model dependence. (e.g. GRB\,060605 and GRB\,080413B mentioned in \citealt{Li_2012} and GRB\,220813A in \citealt{GRB220831A}). Among the studies on optical afterglow of GRBs (e.g., \citealt{Dainotti_2022,Ronchini_2023,Li_2026}), GRB\,070707 \citep{GRB070707} and GRB\,180618A \citep{GRB180618A} stand out as events displaying an uncontaminated optical internal plateau ($\alpha<0.7$ \footnote{throughout the paper the convention $F_{\nu}\propto t^{-\alpha}{\nu}^{-\beta}$ is adopted}) and subsequent steep decay ($\alpha>3$).

During the past two decades, approximately 30 extragalactic fast X-ray transients (eFXTs) have been identified in archival Chandra and XMM-Newton data (e.g., \citealt{Bauer2017,JQV22,JQV23}), the lack of comprehensive multi-wavelength coverage and dedicated follow-up observations hinders definitive classification. 
Launched on January 9, 2024, the Einstein Probe (EP) has revolutionized this field. Using lobster-eye optics, it monitors the soft X-ray sky with unprecedented sensitivity (about 3600 square degrees field of view) and has already detected over 100 transients. While many have been detected both in X-ray and gamma-ray
\citep[e. g.,][]{EP240315a1},
there is a significant proportion of the transients detected by EP have no contemporaneous gamma-ray detection. A subset is classified as GRBs based on their afterglow properties
\citep[e. g.,][]{EP250207b}.
Some are associated with core-collapse supernovae
\citep[e. g.,][]{EP240414a6}.
Nevertheless, there are still some eFXTs remaining of unknown origin, such as EP240408a \citep{EP240408a}, and EP241021a \citep{EP241021a}.

In this Letter, we analyze the eFXT EP251023a detected by EP/WXT (0.5–-4 keV, \citealt{EPhandbook}), which shows an achromatic plateau and steep decay similar to the magnetar associated eFXT CDF-S XT2. Distinctly, EP251023a exhibits prompt soft X-ray emission and an optical counterpart, which were absent in CDF-S XT2 observations. The observed data collected from several facilities are introduced in Section \ref{sec:data}. Section \ref{sec:result} presents our analysis of the prompt emission and the afterglow. In Section \ref{sec:discuss}, we discuss the physics origin of EP251023a and summarize our work. The conventional cosmological model we adopted is as follows: H$_0 = 69.6\,\rm{km}\,\rm{s}^{-1}\,\rm{Mpc}^{-1},\,\Omega_M = 0.286,\,\Omega_{\Lambda} = 0.714$ \citep{Cosmic_param}.

\begin{figure}[htbp!]
\includegraphics[width=0.45\textwidth, keepaspectratio]{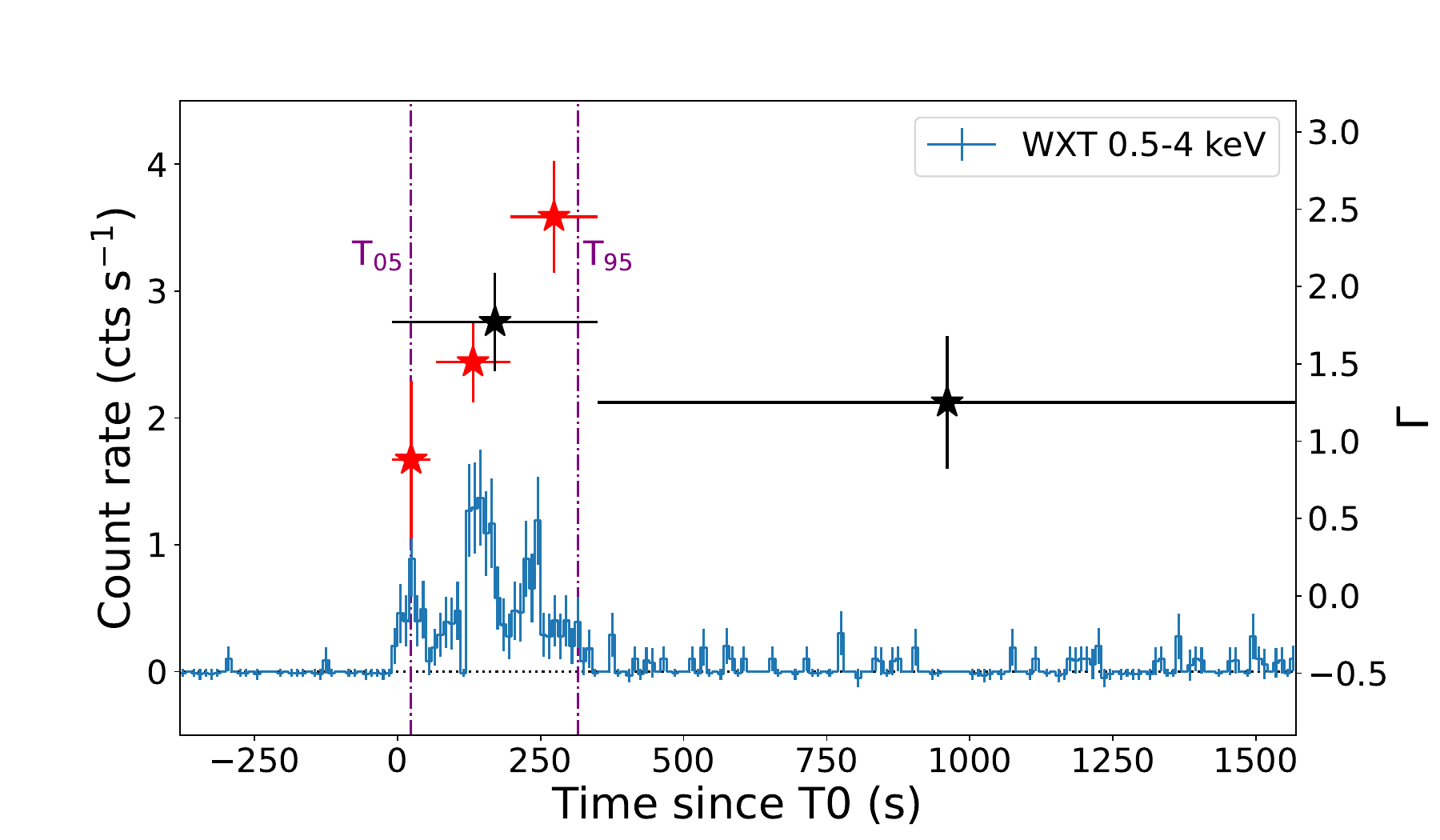}
\caption{EP/WXT light curve of EP251023a. Purple dash-dotted lines mark the 5 and 95 cumulative fluence levels of the main emission. The black stars denote the average spectral indices of the two epochs, while the red stars illustrate the spectral evolution during the main emission phase.
\label{WXT_LC}}
\end{figure}
\section{Observations and data reduction} \label{sec:data}
EP251023a was detected by EP/WXT on 2025-10-23 at 02:29:28 UTC ($T_0$), triggering a transient alert at $T_0 + 101.7$ s upon sufficient signal accumulation. No automatic EP/FXT (0.3–-10 keV; \citealt{FXT}) follow-up was triggered due to ongoing instrument calibration. 
EP/FXT performed three follow-up observations at $T_0 + 0.25$, 1.11, and 3.58 days, detecting the source in the first two epochs but yielding only an upper limit in the last. EP251023a reached an unabsorbed peak flux of $(3.77 \pm 1.34) \times 10^{-9}\,\rm{erg\,s^{-1}\,cm^{-2}}$ (0.5–-4~keV). It is located at R.A.~=~127.1343 deg, Dec.~=~20.8649 deg (J2000) with a 90$\%$ confidence radius of  10$\arcsec$. The light curve of WXT is shown in Fig.~\ref{WXT_LC}. Spectral analysis was performed with \texttt{Xspec v12.14.0h} for EP data; details can be seen in Section \ref{sec:result}. The results of the EP observations are presented in Table \ref{Xray_result}.

Both Swift/BAT and Fermi/GBM were Earth-occulted during the prompt emission, as confirmed by the Swift/GUANO\footnote{\url{https://guano.swift.psu.edu/}} and Fermi/GBM public data\footnote{\url{https://heasarc.gsfc.nasa.gov/FTP/fermi/data/gbm/daily/}}.
At the time of EP251023a, Konus--Wind (KW) was continuously monitoring the entire sky in three energy bands: G1 (18--74 keV), G2 (74--306 keV) and G3 (306--1212 keV), and there is no significant rate increase in the KW data, with an upper limit of $3\times10^{-7} \rm{erg\,s^{-1}\,cm^{-2}}$ in 10--1000 keV at 2.944 s timescale.

\begin{figure}[htbp!]
\includegraphics[width=0.45\textwidth, keepaspectratio]{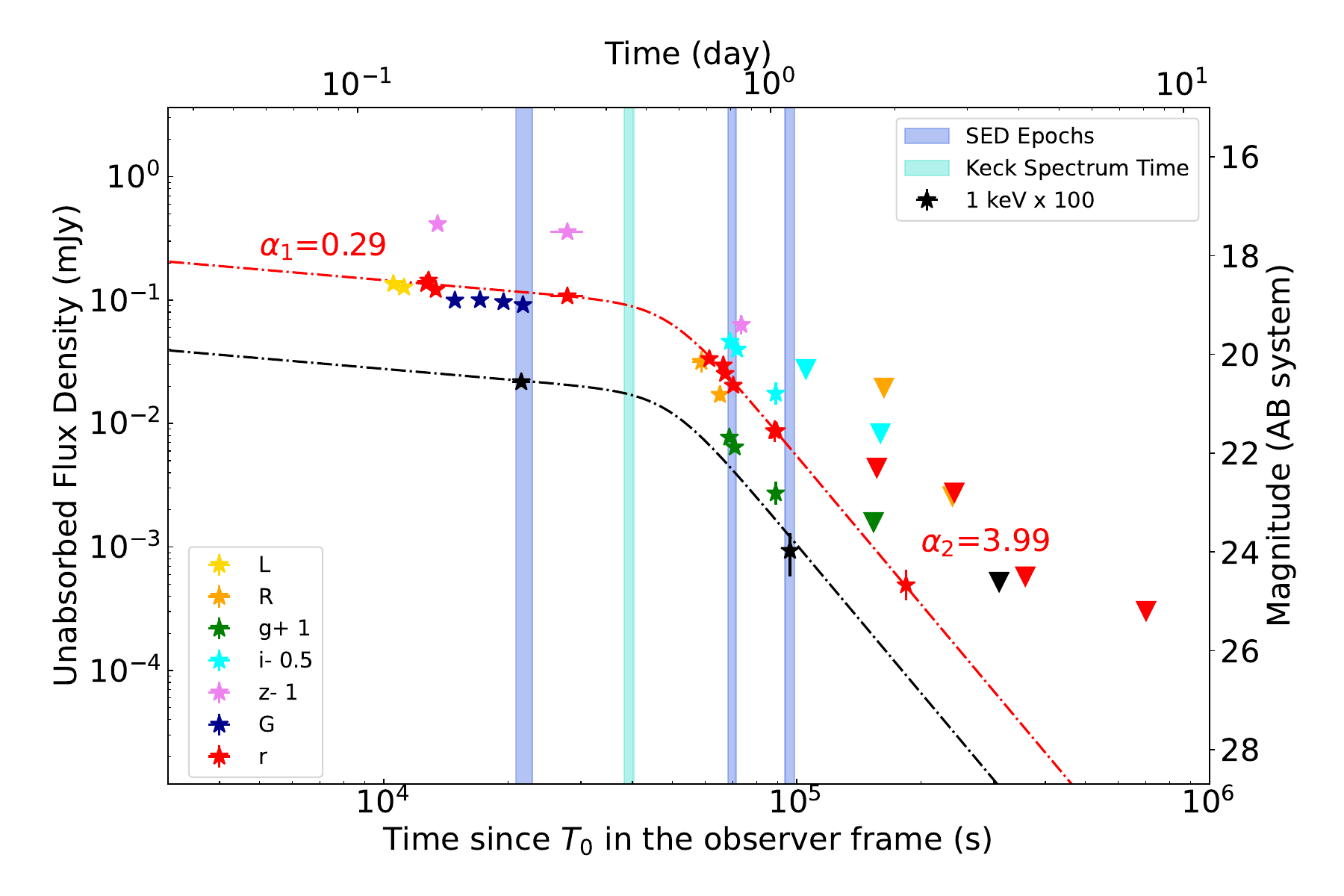}
\caption{Afterglow light curve of EP251023a, showing optical AB magnitudes and 1 keV unabsorbed EP/FXT flux densities. Stars and inverted triangles denote detections and upper limits, respectively. Blue and cyan vertical lines mark the SED and Keck spectrum epochs. The $r$-band data are fitted with a broken power law (red dash-dotted), while the black dash-dotted line shows this model rescaled to the first X-ray point.
\label{totallc}}
\end{figure}

The optical counterpart of EP251023a was located at (J2000) R.A.~=~$08^{\rm hr}28^{\rm m}32.21^{\rm s}$, Dec.~=~$+20^\circ 51' 51.51''$. The celestial location of the burst is shown in Fig.~\ref{locimg}. 
Our follow-up campaign utilized a combination of publicly available GCN\footnote{\url{https://gcn.nasa.gov/circulars}} data and new observations from a suite of ground-based facilities. Details of the filters used with the telescopes, along with the photometric results, are presented in Table \ref{tab:optical_result} and shown in Fig.~\ref{totallc}. 
The optical spectrum of EP251023a was obtained with the Low Resolution Imaging Spectrometer (LRIS) mounted on Keck telescope at about 10.9 hours. Fig.~\ref{optspec} shows the high S/N spectral region (3500--9200\r{A}).
The redshift of EP251023 is identified as $z = 2.232 \pm 0.001$ through numerous strong absorption lines, particularly fine-structure lines. The ﬁt to the strong Ly$\alpha$ absorption line yields a column density of log~($N_{\rm HI}/cm^{-2}$) = $19.5 \pm 0.1$, which is relatively lower than those typically found in GRB host galaxies (see Fig.~\ref{NH_fig}).

\section{Results} \label{sec:result}
In Fig.~\ref{WXT_LC}, we present the light curve of EP251023a with EP/WXT data at 0.5--4 keV with 10~s bin size. The main emission (0--380 s since $T_0$) shows  multiple pulses at 0.5--4 keV with $T_{90} = 292\,\pm\,19\,\rm{s}$. It has an average spectral index of $1.77 \pm 0.32$ with the absorbed power-law model, which yields an average unabsorbed flux in the 0.5--4 keV energy band of $8.74_{-0.95}^{+1.06}\times10^{-10} \rm erg\,s^{-1}\,cm^{-2}$ and a corresponding luminosity of $3.41_{-0.37}^{+0.42}\times10^{49} \rm erg\,s^{-1}$ at $z=2.232$.  The total emission was detected up to $T_0+1571$ s, when the observations ceased and no subsequent data were obtained. The long-lasting emission is well described by an absorbed power-law model with an average spectral index of $1.25 \pm 0.43$, which yields an average unabsorbed flux in the 0.5--4 keV energy band of $6.78_{-1.80}^{+0.91}\times10^{-11} \rm erg\,s^{-1}\,cm^{-2}$ and a corresponding luminosity of $2.66_{-0.71}^{+0.35}\times10^{48} \rm erg\,s^{-1}$.
The soft prompt spectrum ($\Gamma=1.77$) prevents the peak energy $E_{\rm peak}$ from being well constrained by the KW upper limit. However, during the first two intervals (0--57 s and 57--197 s), the spectrum hardens, allowing us to constrain $E_{\rm peak}$ using the KW non-detection. By assuming that the power-law model evolves into a cutoff power-law model as the energy reaches higher, we can derive the upper limits of $E_{\rm peak}$ are 142 keV and 70 keV for the two epochs, respectively. We therefore derive upper limits on $E_{\gamma,\rm{iso}}$ in the rest frame 1--10000 keV for the three phases separately and sum them to obtain a conservative total $E_{\gamma,\rm{iso}}$ upper limit of $5.7 \times 10^{52}$ erg.  

The X-ray and optical data at the afterglow phases are illustrated in Fig.~\ref{totallc}. We fit the $r$-band data with a smoothly broken power-law (SBPL) function $F=F_1[(\frac{t}{t_b})^{\omega \alpha_1}+(\frac{t}{t_b})^{\omega \alpha_2}]^{-\frac{1}{\omega}}$,
where $F_1$ is the flux at the break time ($t_b$), $\alpha_1$ and $\alpha_2$ are the afterglow flux decay indices before and after $t_b$, respectively. $\omega$ quantifies the sharpness of the break—higher values indicate a more abrupt transition. We found that the $r$-band data can be well fitted with $\omega =2$. The data have $\alpha_1=0.29 \pm 0.04$, $\alpha_2=3.99 \pm 0.23$ and break time $t_b=49324 \pm 1566$ s. A line parallel to the optical broken power-law, anchored at the first X-ray data point, is consistent with the X-ray detections. The uniform steep decay across optical bands reveals an achromatic break and subsequent steep decline, which is highly unusual for transient afterglows.

\begin{figure}[htbp!]
\includegraphics[width=0.45\textwidth, keepaspectratio]{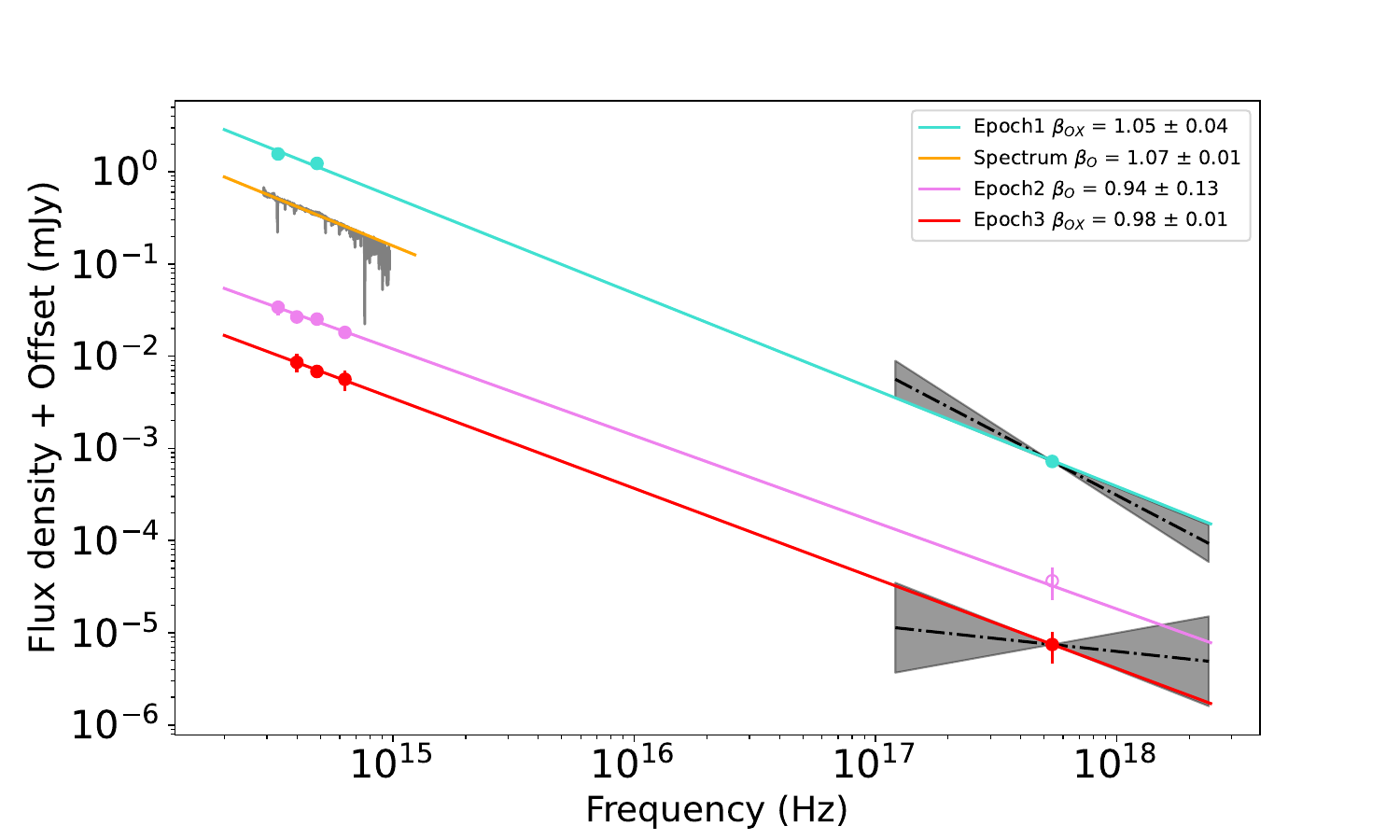}
\caption{SED of EP251023a. Four datasets are vertically offset and ordered by decreasing flux (earliest to latest). The grey line shows the optical spectrum, and solid lines represent best-fit models for SED.Black dash-dotted lines and grey shaded regions indicate individual X-ray fits and their 1-$\sigma$ uncertainties ($\beta_{X,1}=1.37 \pm 0.31$ and $\beta_{X,2}=0.28 \pm 0.73$). The violet open circle denotes the extrapolated X-ray flux (excluded from the fit).
\label{SED}}
\end{figure}

To probe afterglow evolution, we performed spectral energy distribution (SED) analyses at three epochs (blue vertical lines in Fig.~\ref{totallc}) and included the Galactic extinction-corrected Keck spectrum, as illustrated in Fig.~\ref{SED}. The spectral index $\beta$ remains consistent within uncertainties throughout the afterglow phase. The optical data at about 70 ks (indicated by the second blue vertical line in Fig.~\ref{totallc}) yield a spectral index $\beta_{O}=0.94 \pm 0.13$, and the extrapolated X-ray flux density is consistent with the best-fit model (shown as the open violet circle in Fig.~\ref{SED}). Compared to the individual X-ray spectral fit, the joint derived $\beta_{OX}$ is consistent with the $\beta_{X}$, indicating a single power-law spectrum that continuously extends from optical to X-ray frequencies.

\section{Discussion and conclusions}\label{sec:discuss}
The extremely steep post-break decay ($\alpha_2=3.99$ at $t>0.57$ days) cannot be explained by a jet break or a steeply declining ambient density (see Appendix~\ref{sec:origin appendix}). The millisecond magnetar model has been widely applied to the X-ray data of some GRBs and eFXTs, and it can also be applied to interpret the optical afterglows of some sources. As noted in Section~\ref{intro}, GRB\,070707 and GRB\,180618A exhibit an uncontaminated optical internal plateau ($\alpha_1<0.7$) and subsequent steep decay ($\alpha_2>3$). GRB\,180618A is omitted from the comparative sample due to its distinct chromatic and thermal nature, which contrasts with the achromatic afterglows of EP251023a.
Despite some theoretical studies attempting to explain the emission mechanisms of magnetars (e.g., \citealt{Mao_2010,Metzger_2014,Strang_2019}), the underlying physics is not yet fully understood.
The broadband spectra of EP251023a indicate a single power-law spectrum that continuously extends from optical to X-ray frequencies with the spectral index $\beta$ derived from SEDs (i.e. $\beta=1.05$ for the plateau). Thus, we assume a magnetar flux density spectrum following $F_{\nu}\propto {\nu}^{-\beta}$, spanning from $1.4\times10^{14}$ Hz to $2.4\times10^{18}$ Hz (i.e. near-infrared to 10 keV). We utilize this flux density spectrum to derive the luminosity of EP251023a and GRB\,070707, whereas for CDF-S XT2, the luminosity is derived exclusively from soft X-rays given the absence of optical data.

We interpret the observed plateau and subsequent decay within the framework of a millisecond magnetar central engine. In this scenario, the isotropic luminosity of the plateau ($L_{\rm em}$) is powered by the spin-down of a highly magnetized neutron star. Given that no afterglow component was detected for EP251023a throughout the observation period, we assume the kinetic energy contribution is negligible and adopt a radiation efficiency of $\eta \approx 1$. The detailed derivation of the magnetar parameters is provided in Appendix~\ref{sec:origin appendix}. Briefly, in the electromagnetic spin-down-dominated regime, the spin-down luminosity evolves as $L_{\rm em}(t)\propto (1+\frac{t}{\tau})^{-2}$, where $\tau$ is the characteristic spin-down timescale. The observed slope for EP251023a is significantly steeper than the predicted slope of $\alpha \approx 2$, indicating an abrupt cessation of energy injection, likely due to the collapse of a supra-massive magnetar into a black hole. Consequently, the observed break time represents the onset of collapse, implying that the intrinsic spin-down timescale $\tau$ exceeds the observed break time, i.e. $\tau > t_b/(1+z)$.
We fitted the light curve of EP251023a in the plateau stage with Equation \ref{eq5} by using \texttt{REDBACK} \citep{redback}.
From the fit we derived $L_{\rm em}=(2.59\pm0.06)\times 10^{47} \,\rm erg\,s^{-1}$ and $\tau > 15261$ s based on the break time. Isotropic emission is a reasonable assumption for relatively powerful magnetar winds, magnetar-driven outflows following a neutron star merger or an accretion-induced collapse \citep{Bucciantini2011} are unlikely to be efficiently collimated, as there is only a limited amount of surrounding material to confine them. With $f_b=1$,  we derived $P_0<2.27$~ms and $B_p<8.33\times10^{14}$~G for EP251023a. However, the isotropic wind scenario may not hold for the type II GRB magnetars, whose winds are expected to be collimated. As examined in the Swift GRB sample of \citet{lv2014}, most type II GRBs are inconsistent with the isotropic wind scenario. As the physical origin of EP251023a remains unconstrained by current data, we employ a typical jet opening angle of 0.1 rad to apply a beaming correction. With $f_b=0.005$, we derived $P_0<32.15$~ms and $B_p<1.18\times10^{16}$~G for EP251023a.

\begin{figure}[htbp!]
\includegraphics[width=0.45\textwidth, keepaspectratio]{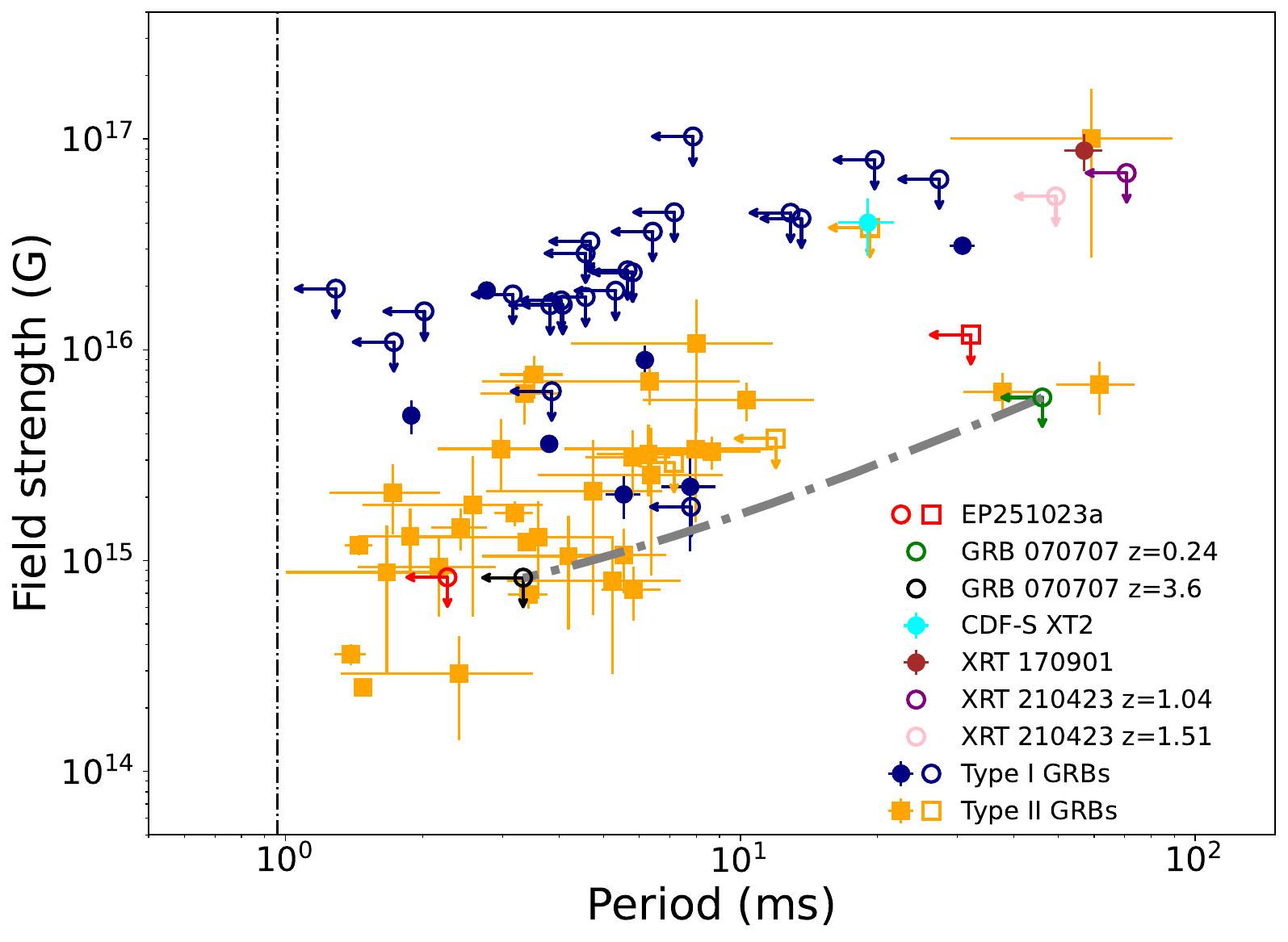}
\caption{Inferred magnetar parameters: $P_0$ vs. $B_p$. Circles and squares denote the isotropic wind ($f_b=1$) and beaming-corrected scenarios, respectively. Sources with only lower limits on $\tau$ are shown as open symbols, and arrows indicate upper limits of $P_0$ and $B_p$. Blue and orange symbols correspond to type I and type II GRBs. The vertical dash-dotted black line represents the neutron star breakup spin-period. Grey dash-dotted lines indicate the redshift evolution of GRB\,070707 from $z=0.24$ (green) to $z=3.6$ (black).
\label{Magnetar compare}}
\end{figure}

For comparison, we also consider other transients associated with millisecond magnetars. CDF-S XT2, XRT\,170901, and XRT\,210423 are eFXTs previously interpreted as magnetar spin-down emission events, while GRB\,070707 shows optical behavior similar to that of EP251023a. Compared with those eFXTs, EP251023a is characterized by a substantially longer plateau duration and the presence of additional prompt emission. Its X-ray spectrum also shows no significant evolution across the break, resembling XRT\,170901 and XRT\,210423 but differing from the spectral softening observed in CDF-S XT2. Together with GRB\,070707, these two sources share a common long-duration plateau. However, they exhibit distinct pre-break broadband SED properties: the X-ray emission of GRB\,070707 is notably brighter than its optical emission, whereas EP251023a does not display such a significant discrepancy. A detailed comparison, including the derivation of the magnetar parameters for these transients, is presented in Appendix~\ref{sec:comparison appendix}. Given that X-ray plateaus followed by steep decays are commonly observed in GRBs, we compare the magnetar parameters of EP251023a with those of the type I and type II Swift GRB samples (\citealt{Rowlinson_2013,lv2014,lv2015}) based exclusively on X-ray observations in Fig.~\ref{Magnetar compare}. The $\tau$ values of some sources are treated as lower limits in our analysis, given that their post-break decay slopes are steeper than 3. These results are illustrated in Fig.~\ref{Magnetar compare}.

In conclusion, EP251023a is an eFXT with a redshift of $z = 2.232\,\pm\,0.001$ detected solely by EP/WXT in the soft X-ray band, and KW does not detect a gamma-ray counterpart. The optical light curves show an achromatic plateau followed by an extremely steep decay with a slope of 3.99 after the break time of about 49 ks.
Such light curve behavior, frequently seen in X-rays but rare in the optical band for some GRBs, suggests a rapidly spinning millisecond magnetar as the central engine. A single power-law decay in the broadband SEDs indicates that the emission extends continuously from the optical to X-ray bands. Given this spectrum, the ratio of the r-band to X-ray luminosity for EP251023a is $\sim 0.1$.  The steep post-break decay of EP251023a implies a magnetar collapse prior to substantial spin-down, setting a lower limit on the spin-down timescale $\tau$ from the observed break time. Under the isotropic wind scenario ($f_b=1$), we obtain $P_0<2.27$~ms and $B_p<8.33\times10^{14}$~G for EP251023a. Under the potential jet collimation assumption, with a beaming factor of $f_b=0.005$, the relaxed magnetar parameters are $P_0<32.15$~ms and $B_p<1.18\times10^{16}$~G. Compared with other eFXTs associated with magnetar spin-down, EP251023a is distinguished by its additional prompt emission and significantly longer plateau, while its X-ray spectrum remains stable throughout the break, similar to XRT\,170901 and XRT\,210423 but unlike the spectral softening observed in CDF-S XT2. Together with GRB\,070707, which also exhibits an optical plateau followed by a very steep decay with little evidence of external-shock emission, EP251023a shares a similar long-duration plateau, but displays different broadband SED properties. The sample of magnetar-powered optical plateaus remains sparse. However, growing detections will help distinguish magnetar internal emissions with optical counterparts from X-ray-only events. Historically, fewer than ten magnetar-associated eFXTs have been identified, mostly lacking redshifts due to delayed archival mining. The extremely steep decay of EP251023a suggests a magnetar-powered origin for its optical plateau. Future EP detections will significantly expand the sample of magnetar-associated eFXTs, enabling us to determine whether magnetars in eFXTs differ from those in GRBs.



%

\bibliographystyle{bibtex/aa}
\bibliography{bibtex/aa}

@article{SF2011,
doi = {10.1088/0004-637X/737/2/103},
url = {https://dx.doi.org/10.1088/0004-637X/737/2/103},
year = {2011},
month = {aug},
publisher = {The American Astronomical Society},
volume = {737},
number = {2},
pages = {103},
author = {Schlafly, Edward F. and Finkbeiner, Douglas P.},
title = {MEASURING REDDENING WITH SLOAN DIGITAL SKY SURVEY STELLAR SPECTRA AND RECALIBRATING SFD},
journal = {The Astrophysical Journal}
}

@INPROCEEDINGS{FXT,
       author = {{Chen}, Yong and {Cui}, WeiWei and {Han}, DaWei and {Wang}, Juan and {Yang}, YanJi and {Wang}, YuSa and {Li}, Wei and {Ma}, Jia and {Xu}, YuPeng and {Lu}, FangJun and {Chen}, HouLei and {Tang}, QingJun and {Yuan}, Weimin and {Friedrich}, Peter and {Meidinger}, Norbert and {Keil}, Isabell and {Burwitz}, Vadim and {Eder}, Josef and {Hartmann}, Katinka and {Nandra}, Kirpal and {Keereman}, Arnoud and {Santovincenzo}, Andrea and {Vernani}, Dervis and {Bianucci}, Giovanni and {Valsecchi}, Giuseppe and {Wang}, Bo and {Wang}, LangPing and {Wang}, DianLong and {Li}, Duo and {Sheng}, LiZhi and {Qiang}, PengFei and {Shi}, RongRong and {Chao}, XiangYu and {Song}, Zeyu and {Zhang}, Ziliang and {Huo}, Jia and {Wang}, Hao and {Cong}, Min and {Yang}, XiongTao and {Hou}, Dongjie and {Zhao}, XiaoFan and {Zhao}, ZiJian and {Chen}, TianXiang and {Li}, MaoShun and {Zhang}, Tong and {Luo}, LaiDan and {Xu}, JingJing and {Li}, Gang and {Zhang}, Qian and {Bi}, XiYan and {Zhu}, YuXuan and {Yu}, Nian and {Chen}, Can and {Lv}, ZhongHua and {Lu}, Bing and {Zhang}, JiaWei},
        title = "{Status of the follow-up x-ray telescope onboard the Einstein Probe satellite}",
    booktitle = {Society of Photo-Optical Instrumentation Engineers (SPIE) Conference Series},
         year = 2021,
       editor = {{den Herder}, Jan-Willem A. and {Nikzad}, Shouleh and {Nakazawa}, Kazuhiro},
       series = {Society of Photo-Optical Instrumentation Engineers (SPIE) Conference Series},
       volume = {11444},
        month = jan,
          eid = {114445B},
        pages = {114445B},
          doi = {10.1117/12.2562311},
       adsurl = {https://ui.adsabs.harvard.edu/abs/2021SPIE11444E..5BC}
}

@Inbook{EPhandbook,
author="Yuan, Weimin
and Zhang, Chen
and Chen, Yong
and Ling, Zhixing",
editor="Bambi, Cosimo
and Santangelo, Andrea",
title="The Einstein Probe Mission",
bookTitle="Handbook of X-ray and Gamma-ray Astrophysics",
year="2022",
publisher="Springer Nature Singapore",
address="Singapore",
pages="1--30",
isbn="978-981-16-4544-0",
doi="10.1007/978-981-16-4544-0_151-1",
url="https://doi.org/10.1007/978-981-16-4544-0_151-1"
}

@ARTICLE{GOTOgcn,
       author = {{Wortley}, M.~E. and {O'Neill}, D. and {Ramsay}, G. and {Gompertz}, B.~P. and {Godson}, B. and {Starling}, R. and {Ackley}, K. and {Dyer}, M.~J. and {Lyman}, J. and {Ulaczyk}, K. and {Steeghs}, D. and {Galloway}, D.~K. and {Dhillon}, V. and {O'Brien}, P. and {Noysena}, K. and {Kotak}, R. and {Breton}, R.~P. and {Nuttall}, L.~K. and {Casares}, J. and {GOTO Collaboration}},
        title = "{EP/WXT 01709247295: GOTO optical afterglow candidate.}",
      journal = {GRB Coordinates Network},
         year = 2025,
        month = oct,
       volume = {42387},
        pages = {1},
       adsurl = {https://ui.adsabs.harvard.edu/abs/2025GCN.42387....1W}
}

@ARTICLE{2019MNRAS.483.5380T,
       author = {{Tanvir}, N.~R. and {Fynbo}, J.~P.~U. and {de Ugarte Postigo}, A. and {Japelj}, J. and {Wiersema}, K. and {Malesani}, D. and {Perley}, D.~A. and {Levan}, A.~J. and {Selsing}, J. and {Cenko}, S.~B. and {Kann}, D.~A. and {Milvang-Jensen}, B. and {Berger}, E. and {Cano}, Z. and {Chornock}, R. and {Covino}, S. and {Cucchiara}, A. and {D'Elia}, V. and {Gargiulo}, A. and {Goldoni}, P. and {Gomboc}, A. and {Heintz}, K.~E. and {Hjorth}, J. and {Izzo}, L. and {Jakobsson}, P. and {Kaper}, L. and {Kr{\"u}hler}, T. and {Laskar}, T. and {Myers}, M. and {Piranomonte}, S. and {Pugliese}, G. and {Rossi}, A. and {S{\'a}nchez-Ram{\'\i}rez}, R. and {Schulze}, S. and {Sparre}, M. and {Stanway}, E.~R. and {Tagliaferri}, G. and {Th{\"o}ne}, C.~C. and {Vergani}, S. and {Vreeswijk}, P.~M. and {Wijers}, R.~A.~M.~J. and {Watson}, D. and {Xu}, D.},
        title = "{The fraction of ionizing radiation from massive stars that escapes to the intergalactic medium}",
      journal = {\mnras},
         year = 2019,
        month = mar,
       volume = {483},
       number = {4},
        pages = {5380-5408},
          doi = {10.1093/mnras/sty3460},
archivePrefix = {arXiv},
       eprint = {1805.07318},
 primaryClass = {astro-ph.GA},
       adsurl = {https://ui.adsabs.harvard.edu/abs/2019MNRAS.483.5380T}
}

@ARTICLE{1999ApJ...525..737R,
       author = {{Rhoads}, James E.},
        title = "{The Dynamics and Light Curves of Beamed Gamma-Ray Burst Afterglows}",
      journal = {\apj},
         year = 1999,
        month = nov,
       volume = {525},
       number = {2},
        pages = {737-749},
          doi = {10.1086/307907},
archivePrefix = {arXiv},
       eprint = {astro-ph/9903399},
 primaryClass = {astro-ph},
       adsurl = {https://ui.adsabs.harvard.edu/abs/1999ApJ...525..737R}
}

@article{Zhang_2001,
doi = {10.1086/320255},
url = {https://doi.org/10.1086/320255},
year = {2001},
month = {apr},
publisher = {},
volume = {552},
number = {1},
pages = {L35},
author = {Zhang, Bing and Mészáros, Peter},
title = {Gamma-Ray Burst Afterglow with Continuous Energy Injection: Signature of a Highly Magnetized Millisecond Pulsar},
journal = {The Astrophysical Journal}
}

@ARTICLE{Cosmic_param,
       author = {{Bennett}, C.~L. and {Larson}, D. and {Weiland}, J.~L. and {Hinshaw}, G.},
        title = "{The 1\% Concordance Hubble Constant}",
      journal = {\apj},
         year = 2014,
        month = oct,
       volume = {794},
       number = {2},
          eid = {135},
        pages = {135},
          doi = {10.1088/0004-637X/794/2/135},
archivePrefix = {arXiv},
       eprint = {1406.1718},
 primaryClass = {astro-ph.CO},
       adsurl = {https://ui.adsabs.harvard.edu/abs/2014ApJ...794..135B}
}

@ARTICLE{density_jump,
       author = {{Nakar}, Ehud and {Granot}, Jonathan},
        title = "{Smooth light curves from a bumpy ride: relativistic blast wave encounters a density jump}",
      journal = {\mnras},
         year = 2007,
        month = oct,
       volume = {380},
       number = {4},
        pages = {1744-1760},
          doi = {10.1111/j.1365-2966.2007.12245.x},
archivePrefix = {arXiv},
       eprint = {astro-ph/0606011},
 primaryClass = {astro-ph},
       adsurl = {https://ui.adsabs.harvard.edu/abs/2007MNRAS.380.1744N}
}

@ARTICLE{GRB070707,
       author = {{Piranomonte}, S. and {D'Avanzo}, P. and {Covino}, S. and {Antonelli}, L.~A. and {Beardmore}, A.~P. and {Campana}, S. and {Chincarini}, G. and {D'Elia}, V. and {Della Valle}, M. and {Fiore}, F. and {Fugazza}, D. and {Guetta}, D. and {Guidorzi}, C. and {Israel}, G.~L. and {Lazzati}, D. and {Malesani}, D. and {Parsons}, A.~M. and {Perna}, R. and {Stella}, L. and {Tagliaferri}, G. and {Vergani}, S.~D.},
        title = "{The short GRB 070707 afterglow and its very faint host galaxy}",
      journal = {\aap},
         year = 2008,
        month = nov,
       volume = {491},
       number = {1},
        pages = {183-188},
          doi = {10.1051/0004-6361:200810547},
archivePrefix = {arXiv},
       eprint = {0807.1348},
 primaryClass = {astro-ph},
       adsurl = {https://ui.adsabs.harvard.edu/abs/2008A&A...491..183P}
}

@ARTICLE{Xue_nature,
       author = {{Xue}, Y.~Q. and {Zheng}, X.~C. and {Li}, Y. and {Brandt}, W.~N. and {Zhang}, B. and {Luo}, B. and {Zhang}, B.-B. and {Bauer}, F.~E. and {Sun}, H. and {Lehmer}, B.~D. and {Wu}, X.-F. and {Yang}, G. and {Kong}, X. and {Li}, J.~Y. and {Sun}, M.~Y. and {Wang}, J.-X. and {Vito}, F.},
        title = "{A magnetar-powered X-ray transient as the aftermath of a binary neutron-star merger}",
      journal = {\nat},
         year = 2019,
        month = apr,
       volume = {568},
       number = {7751},
        pages = {198-201},
          doi = {10.1038/s41586-019-1079-5},
archivePrefix = {arXiv},
       eprint = {1904.05368},
 primaryClass = {astro-ph.HE},
       adsurl = {https://ui.adsabs.harvard.edu/abs/2019Natur.568..198X}
}

@ARTICLE{EP240315a1,
       author = {{Liu}, Y. and {Sun}, H. and {Xu}, D. and {Svinkin}, D.~S. and {Delaunay}, J. and {Tanvir}, N.~R. and {Gao}, H. and {Zhang}, C. and {Chen}, Y. and {Wu}, X.-F. and {Zhang}, B. and {Yuan}, W. and {An}, J. and {Bruni}, G. and {Frederiks}, D.~D. and {Ghirlanda}, G. and {Hu}, J.-W. and {Li}, A. and {Li}, C.-K. and {Li}, J.-D. and {Malesani}, D.~B. and {Piro}, L. and {Raman}, G. and {Ricci}, R. and {Troja}, E. and {Vergani}, S.~D. and {Wu}, Q.-Y. and {Yang}, J. and {Zhang}, B.-B. and {Zhu}, Z.-P. and {de Ugarte Postigo}, A. and {Demin}, A.~G. and {Dobie}, D. and {Fan}, Z. and {Fu}, S.-Y. and {Fynbo}, J.~P.~U. and {Geng}, J.-J. and {Gianfagna}, G. and {Hu}, Y.-D. and {Huang}, Y.-F. and {Jiang}, S.-Q. and {Jonker}, P.~G. and {Julakanti}, Y. and {Kennea}, J.~A. and {Kokomov}, A.~A. and {Kuulkers}, E. and {Lei}, W.-H. and {Leung}, J.~K. and {Levan}, A.~J. and {Li}, D.-Y. and {Li}, Y. and {Littlefair}, S.~P. and {Liu}, X. and {Lysenko}, A.~L. and {Ma}, Y.-N. and {Martin-Carrillo}, A. and {O'Brien}, P. and {Parsotan}, T. and {Quirola-V{\'a}squez}, J. and {Ridnaia}, A.~V. and {Ronchini}, S. and {Rossi}, A. and {Mata-S{\'a}nchez}, D. and {Schneider}, B. and {Shen}, R.-F. and {Thakur}, A.~L. and {Tohuvavohu}, A. and {Torres}, M.~A.~P. and {Tsvetkova}, A.~E. and {Ulanov}, M.~V. and {Wei}, J.-J. and {Xiao}, D. and {Yin}, Y.-H.~I. and {Bai}, M. and {Burwitz}, V. and {Cai}, Z.-M. and {Chen}, F.-S. and {Chen}, H.-L. and {Chen}, T.-X. and {Chen}, W. and {Chen}, Y.-F. and {Chen}, Y.-H. and {Cheng}, H.-Q. and {Cordier}, B. and {Cui}, C.-Z. and {Cui}, W.-W. and {Dai}, Y.-F. and {Dai}, Z.-G. and {Eder}, J. and {Eyles-Ferris}, R.~A.~J. and {Fan}, D.-W. and {Feldman}, C. and {Feng}, H. and {Feng}, Z. and {Friedrich}, P. and {Gao}, X. and {Gonzalez}, J.-F. and {Guan}, J. and {Han}, D.-W. and {Han}, J. and {Hou}, D.-J. and {Hu}, H.-B. and {Hu}, T. and {Huang}, M.-H. and {Huo}, J. and {Hutchinson}, I. and {Ji}, Z. and {Jia}, S.-M. and {Jia}, Z.-Q. and {Jiang}, B.-W. and {Jin}, C.-C. and {Jin}, G. and {Jin}, J.-J. and {Keereman}, A. and {Lerman}, H. and {Li}, J.-F. and {Li}, L.-H. and {Li}, M.-S. and {Li}, W. and {Li}, Z.-D. and {Lian}, T.-Y. and {Liang}, E.-W. and {Ling}, Z.-X. and {Liu}, C.-Z. and {Liu}, H.-Y. and {Liu}, H.-Q. and {Liu}, M.-J. and {Liu}, Y.-R. and {Lu}, F.-J. and {L{\"u}}, H.-J. and {Luo}, L.-D. and {Ma}, F.~L. and {Ma}, J. and {Mao}, J.-R. and {Mao}, X. and {McHugh}, M. and {Meidinger}, N. and {Nandra}, K. and {Osborne}, J.~P. and {Pan}, H.-W. and {Pan}, X. and {Ravasio}, M.~E. and {Rau}, A. and {Rea}, N. and {Rehman}, U. and {Sanders}, J. and {Santovincenzo}, A. and {Song}, L.-M. and {Su}, J. and {Sun}, L.-J. and {Sun}, S.-L. and {Sun}, X.-J. and {Tan}, Y.-Y. and {Tang}, Q.-J. and {Tao}, Y.-H. and {Tong}, J.-Z. and {Wang}, C.-Y. and {Wang}, H. and {Wang}, J. and {Wang}, L. and {Wang}, W.-X. and {Wang}, X.-F. and {Wang}, X.-Y. and {Wang}, Y.-L. and {Wang}, Y.-S. and {Wei}, D.-M. and {Willingale}, R. and {Xiong}, S.-L. and {Xu}, H.-T. and {Xu}, J.-J. and {Xu}, X.-P. and {Xu}, Y.-F. and {Xu}, Z. and {Xue}, C.-B. and {Xue}, Y.-L. and {Yan}, A.-L. and {Yang}, F. and {Yang}, H.-N. and {Yang}, X.-T. and {Yang}, Y.-J. and {Yu}, Y.-W. and {Zhang}, J. and {Zhang}, M. and {Zhang}, S.-N. and {Zhang}, W.-D. and {Zhang}, W.-J. and {Zhang}, Y.-H. and {Zhang}, Z. and {Zhang}, Z. and {Zhang}, Z.-L. and {Zhao}, D.-H. and {Zhao}, H.-S. and {Zhao}, X.-F. and {Zhao}, Z.-J. and {Zhou}, L.-X. and {Zhou}, Y.-L. and {Zhu}, Y.-X. and {Zhu}, Z.-C. and {Zuo}, X.-X.},
        title = "{Soft X-ray prompt emission from the high-redshift gamma-ray burst EP240315a}",
      journal = {Nature Astronomy},
         year = 2025,
        month = apr,
       volume = {9},
        pages = {564-576},
          doi = {10.1038/s41550-024-02449-8},
archivePrefix = {arXiv},
       eprint = {2404.16425},
 primaryClass = {astro-ph.HE},
       adsurl = {https://ui.adsabs.harvard.edu/abs/2025NatAs...9..564L}
}

@ARTICLE{EP240414a6,
       author = {{Sun}, H. and {Li}, W.-X. and {Liu}, L.-D. and {Gao}, H. and {Wang}, X.-F. and {Yuan}, W. and {Zhang}, B. and {Filippenko}, A.~V. and {Xu}, D. and {An}, T. and {Ai}, S. and {Brink}, T.~G. and {Liu}, Y. and {Liu}, Y.-Q. and {Wang}, C.-Y. and {Wu}, Q.-Y. and {Wu}, X.-F. and {Yang}, Y. and {Zhang}, B.-B. and {Zheng}, W.-K. and {Ahumada}, T. and {Dai}, Z.-G. and {Delaunay}, J. and {Elias-Rosa}, N. and {Benetti}, S. and {Fu}, S.-Y. and {Howell}, D.~A. and {Huang}, Y.-F. and {Kasliwal}, M.~M. and {Karambelkar}, V. and {Stein}, R. and {Lei}, W.-H. and {Lian}, T.-Y. and {Peng}, Z.-K. and {Frederiks}, D.~D. and {Ridnaia}, A.~V. and {Svinkin}, D.~S. and {Wang}, X.-Y. and {Wang}, A.-L. and {Wei}, D.-M. and {An}, J. and {Andrews}, M. and {Bai}, J.-M. and {Dai}, C.-Y. and {Ehgamberdiev}, S.~A. and {Fan}, Z. and {Farah}, J. and {Feng}, H.-C. and {Fynbo}, J.~P.~U. and {Guo}, W.-J. and {Guo}, Z. and {Hu}, M.-K. and {Hu}, J.-W. and {Jiang}, S.-Q. and {Jin}, J.-J. and {Li}, A. and {Li}, J.-D. and {Li}, R.-Z. and {Liang}, Y.-F. and {Ling}, Z.-X. and {Liu}, X. and {Mao}, J.-R. and {McCully}, C. and {Mirzaqulov}, D. and {Newsome}, M. and {Padilla Gonzalez}, E. and {Pan}, X. and {Terreran}, G. and {Tinyanont}, S. and {Wang}, B.-T. and {Wang}, L.-Z. and {Wen}, X.-D. and {Xiang}, D.-F. and {Xue}, S.-J. and {Yang}, J. and {Zhu}, Z.-P. and {Cai}, Z.-M. and {Castro-Tirado}, A.~J. and {Chen}, F.-S. and {Chen}, H.-L. and {Chen}, T.-X. and {Chen}, W. and {Chen}, Y.-H. and {Chen}, Y.-F. and {Chen}, Y. and {Cheng}, H.-Q. and {Cordier}, B. and {Cui}, C.-Z. and {Cui}, W.-W. and {Dai}, Y.-F. and {Fan}, D.-W. and {Feng}, H. and {Guan}, J. and {Han}, D.-W. and {Hou}, D.-J. and {Hu}, H.-B. and {Huang}, M.-H. and {Huo}, J. and {Jia}, S.-M. and {Jia}, Z.-Q. and {Jiang}, B.-W. and {Jin}, C.-C. and {Jin}, G. and {Kuulkers}, E. and {Li}, C.-K. and {Li}, D.-Y. and {Li}, J.-F. and {Li}, L.-H. and {Li}, M.-S. and {Li}, W. and {Li}, Z.-D. and {Liu}, C.-Z. and {Liu}, H.-Y. and {Liu}, H.-Q. and {Liu}, M.-J. and {Lu}, F.-J. and {Luo}, L.-D. and {Ma}, J. and {Mao}, X. and {Nandra}, K. and {O'Brien}, P. and {Pan}, H.-W. and {Rau}, A. and {Rea}, N. and {Sanders}, J. and {Song}, L.-M. and {Sun}, S.-L. and {Sun}, X.-J. and {Tan}, Y.-Y. and {Tang}, Q.-J. and {Tao}, Y.-H. and {Wang}, H. and {Wang}, J. and {Wang}, L. and {Wang}, W.-X. and {Wang}, Y.-L. and {Wang}, Y.-S. and {Xiong}, D.-R. and {Xu}, H.-T. and {Xu}, J.-J. and {Xu}, X.-P. and {Xu}, Y.-F. and {Xu}, Z. and {Xue}, C.-B. and {Xue}, Y.-L. and {Yan}, A.-L. and {Yang}, H.-N. and {Yang}, X.-T. and {Yang}, Y.-J. and {Zhang}, C. and {Zhang}, J. and {Zhang}, M. and {Zhang}, S.-N. and {Zhang}, W.-D. and {Zhang}, W.-J. and {Zhang}, Y.-H. and {Zhang}, Z. and {Zhang}, Z. and {Zhang}, Z.-L. and {Zhao}, D.-H. and {Zhao}, H.-S. and {Zhao}, X.-F. and {Zhao}, Z.-J. and {Zhou}, Y.-L. and {Zhu}, Y.-X. and {Zhu}, Z.-C. and {Zou}, H.},
        title = "{A fast X-ray transient from a weak relativistic jet associated with a type Ic-BL supernova}",
      journal = {Nature Astronomy},
         year = 2025,
        month = jul,
       volume = {9},
        pages = {1073-1085},
          doi = {10.1038/s41550-025-02571-1},
archivePrefix = {arXiv},
       eprint = {2410.02315},
 primaryClass = {astro-ph.HE},
       adsurl = {https://ui.adsabs.harvard.edu/abs/2025NatAs...9.1073S}
}

@ARTICLE{EP250207b,
       author = {{Jonker}, P.~G. and {Levan}, A.~J. and {Liu}, Xing and {Xu}, Dong and {Liu}, Yuan and {Xu}, Xinpeng and {Li}, An and {Sarin}, N. and {Tanvir}, N.~R. and {Lamb}, G.~P. and {Ravasio}, M.~E. and {S{\'a}nchez-Sierras}, J. and {Quirola-V{\'a}squez}, J.~A. and {Rayson}, B.~C. and {van Dalen}, J.~N.~D. and {Malesani}, D.~B. and {van Hoof}, A.~P.~C. and {Bauer}, F.~E. and {Chac{\'o}n}, J. and {Smartt}, S.~J. and {Martin-Carrillo}, A. and {Corcoran}, G. and {Cotter}, L. and {Rossi}, A. and {Onori}, F. and {Fraser}, M. and {O'Brien}, P.~T. and {Eyles-Ferris}, R.~A.~J. and {Hjorth}, J. and {Chen}, T.-W. and {Leloudas}, G. and {Tomasella}, L. and {Schulze}, S. and {De Pasquale}, M. and {Carotenuto}, F. and {Bright}, J. and {Wang}, Chenwei and {Xiong}, Shaolin and {Zhang}, Jinpeng and {Xue}, Wangchen and {Liu}, Jiacong and {Li}, Chengkui and {Mata S{\'a}nchez}, D. and {Torres}, M.~A.~P.},
        title = "{EP250207b is not a collapsar fast X-ray transient. Is it due to a binary compact object merger?}",
      journal = {\mnras},
         year = 2026,
        month = jan,
       volume = {545},
       number = {2},
          eid = {staf2021},
        pages = {staf2021},
          doi = {10.1093/mnras/staf2021},
archivePrefix = {arXiv},
       eprint = {2508.13039},
 primaryClass = {astro-ph.HE},
       adsurl = {https://ui.adsabs.harvard.edu/abs/2026MNRAS.545f2021J}
}

@ARTICLE{EP240408a,
       author = {{Zhang}, Wenda and {Yuan}, Weimin and {Ling}, Zhixing and {Chen}, Yong and {Rea}, Nanda and {Rau}, Arne and {Cai}, Zhiming and {Cheng}, Huaqing and {Coti Zelati}, Francesco and {Dai}, Lixin and {Hu}, Jingwei and {Jia}, Shumei and {Jin}, Chichuan and {Li}, Dongyue and {O'Brien}, Paul and {Shen}, Rongfeng and {Shu}, Xinwen and {Sun}, Shengli and {Sun}, Xiaojin and {Wang}, Xiaofeng and {Yang}, Lei and {Zhang}, Bing and {Zhang}, Chen and {Zhang}, Shuang-Nan and {Zhang}, Yonghe and {An}, Jie and {Buckley}, David and {Coleiro}, Alexis and {Cordier}, Bertrand and {Dou}, Liming and {Eyles-Ferris}, Rob and {Fan}, Zhou and {Feng}, Hua and {Fu}, Shaoyu and {Fynbo}, Johan P.~U. and {Galbany}, Lluis and {Jha}, Saurabh W. and {Jiang}, Shuaiqing and {Kong}, Albert and {Kuulkers}, Erik and {Lei}, Weihua and {Li}, Wenxiong and {Liu}, Bifang and {Liu}, Mingjun and {Liu}, Xing and {Liu}, Yuan and {Liu}, Zhu and {Maitra}, Chandreyee and {Marino}, Alessio and {Monageng}, Itumeleng and {Nandra}, Kirpal and {Sanders}, Jeremy and {Soria}, Roberto and {Tao}, Lian and {Wang}, Junfeng and {Wang}, Song and {Wang}, Tinggui and {Wang}, Zhongxiang and {Wu}, Qingwen and {Wu}, Xuefeng and {Xu}, Dong and {Xu}, Yanjun and {Xue}, Suijian and {Xue}, Yongquan and {Zhang}, Zijian and {Zhu}, Zipei and {Zou}, Hu and {Bao}, Congying and {Chen}, Fansheng and {Chen}, Houlei and {Chen}, Tianxiang and {Chen}, Wei and {Chen}, Yehai and {Chen}, Yifan and {Cui}, Chenzhou and {Cui}, Weiwei and {Dai}, Yanfeng and {Fan}, Dongwei and {Guan}, Ju and {Han}, Dawei and {Hou}, Dongjie and {Hu}, Haibo and {Huang}, Maohai and {Huo}, Jia and {Jia}, Zhenqing and {Jiang}, Bowen and {Jin}, Ge and {Li}, Chengkui and {Li}, Junfei and {Li}, Longhui and {Li}, Maoshun and {Li}, Wei and {Li}, Zhengda and {Lian}, Tianying and {Liu}, Congzhan and {Liu}, Heyang and {Liu}, Huaqiu and {Lu}, Fangjun and {Luo}, Laidan and {Ma}, Jia and {Mao}, Xuan and {Pan}, Haiwu and {Pan}, Xin and {Song}, Liming and {Sun}, Hui and {Tan}, Yunyin and {Tang}, Qingjun and {Tao}, Yihan and {Wang}, Hao and {Wang}, Juan and {Wang}, Lei and {Wang}, Wenxin and {Wang}, Yilong and {Wang}, Yusa and {Wu}, Qinyu and {Xu}, Haitao and {Xu}, Jingjing and {Xu}, Xinpeng and {Xu}, Yunfei and {Xu}, Zhao and {Xue}, Changbin and {Xue}, Yulong and {Yan}, Ailiang and {Yang}, Haonan and {Yang}, Xiongtao and {Yang}, Yanji and {Zhang}, Juan and {Zhang}, Mo and {Zhang}, Wenjie and {Zhang}, Zhen and {Zhang}, Zhen and {Zhang}, Ziliang and {Zhao}, Donghua and {Zhao}, Haisheng and {Zhao}, Xiaofan and {Zhao}, Zijian and {Zhou}, Hongyan and {Zhou}, Yilin and {Zhu}, Yuxuan and {Zhu}, Zhencai},
        title = "{Einstein Probe discovery of EP240408a: A peculiar X-ray transient with an intermediate timescale}",
      journal = {Science China Physics, Mechanics, and Astronomy},
         year = 2025,
        month = jan,
       volume = {68},
       number = {1},
          eid = {219511},
        pages = {219511},
          doi = {10.1007/s11433-024-2524-4},
archivePrefix = {arXiv},
       eprint = {2410.21617},
 primaryClass = {astro-ph.HE},
       adsurl = {https://ui.adsabs.harvard.edu/abs/2025SCPMA..6819511Z}
}

@ARTICLE{EP241021a,
       author = {{Shu}, Xinwen and {Yang}, Lei and {Yang}, Haonan and {Xu}, Fan and {Chen}, Jin-Hong and {Eyles-Ferris}, Rob A.~J. and {Dai}, Lixin and {Yu}, Yunwei and {Shen}, Rong-Feng and {Sun}, Luming and {Ding}, Hucheng and {Zheng}, WeiKang and {Jiang}, Ning and {Li}, Wenxiong and {Sun}, Ning-Chen and {Xu}, Dong and {Zhang}, Zhumao and {Jin}, Chichuan and {Rau}, Arne and {Wang}, Tinggui and {Wu}, Xue-feng and {Yuan}, Weimin and {Zhang}, Bing and {Nandra}, Kirpal and {Filippenko}, Alexei V. and {Poidevin}, Fr{\'e}d{\'e}rick and {Soria}, Roberto and {Kumar}, Amit and {Aguado}, David S. and {An}, Fangxia and {An}, Tao and {An}, Jie and {Andrews}, Moira and {Anutarawiramkul}, Rungrit and {Baldini}, Pietro and {Brink}, Thomas G. and {Butpan}, Pathompong and {Cai}, Zhiming and {Castro-Tirado}, Alberto J. and {Cheng}, Huaqing and {Cui}, Weiwei and {Farah}, Joseph and {Fu}, Shaoyu and {Fynbo}, Johan P.~U. and {Gao}, Xing and {Han}, Dawei and {Han}, Xuhui and {Howell}, D. Andrew and {Hu}, Jingwei and {Jiang}, Shuaiqing and {Kumar}, Brajesh and {Lei}, Weihua and {Li}, Dongyue and {Li}, Chengkui and {Liu}, Huaqiu and {Liu}, Xing and {Liu}, Yuan and {Liu}, Xiaowei and {L{\'o}pez-Oramas}, Alicia and {L{\'o}pez Fern{\'a}ndez-Nespral}, David and {Maund}, Justyn R. and {McCully}, Curtis and {Niu}, Zexi and {Newsome}, Megan and {O'Brien}, Paul and {Pan}, Haiwu and {Pan}, Yu and {Padilla Gonzalez}, Estefania and {P{\'e}rez-Fournon}, Ismael and {Silima}, Walter and {Sun}, Hui and {Sun}, Shengli and {Sun}, Xiaojin and {Terreran}, Giacomo and {Tinyanont}, Samaporn and {Wang}, Junxian and {Wang}, Yanan and {Wang}, Yun and {Wiersema}, Klaas and {Xu}, Yunfei and {Xue}, Yongquan and {Yang}, Yi and {Zhang}, Fabao and {Zhang}, Juan and {Zhang}, Pinpin and {Zhang}, Wenda and {Zhang}, Yonghe and {Zhao}, Haisheng and {Zhu}, Zipei and {Xin}, Liping and {Yao}, Zhuheng and {Cordier}, Bertrand and {Wei}, Jianyan and {Qiu}, Yulei and {Daigne}, Fr{\'e}d{\'e}ric},
        title = "{EP241021a: A Months-duration X-Ray Transient with Luminous Optical and Radio Emission}",
      journal = {\apjl},
         year = 2025,
        month = sep,
       volume = {990},
       number = {1},
          eid = {L29},
        pages = {L29},
          doi = {10.3847/2041-8213/adf4cd},
archivePrefix = {arXiv},
       eprint = {2505.07665},
 primaryClass = {astro-ph.HE},
       adsurl = {https://ui.adsabs.harvard.edu/abs/2025ApJ...990L..29S}
}

@ARTICLE{Bauer2017,
       author = {{Bauer}, Franz E. and {Treister}, Ezequiel and {Schawinski}, Kevin and {Schulze}, Steve and {Luo}, Bin and {Alexander}, David M. and {Brandt}, William N. and {Comastri}, Andrea and {Forster}, Francisco and {Gilli}, Roberto and {Kann}, David Alexander and {Maeda}, Keiichi and {Nomoto}, Ken'ichi and {Paolillo}, Maurizio and {Ranalli}, Piero and {Schneider}, Donald P. and {Shemmer}, Ohad and {Tanaka}, Masaomi and {Tolstov}, Alexey and {Tominaga}, Nozomu and {Tozzi}, Paolo and {Vignali}, Cristian and {Wang}, Junxian and {Xue}, Yongquan and {Yang}, Guang},
        title = "{A new, faint population of X-ray transients}",
      journal = {\mnras},
         year = 2017,
        month = jun,
       volume = {467},
       number = {4},
        pages = {4841-4857},
          doi = {10.1093/mnras/stx417},
archivePrefix = {arXiv},
       eprint = {1702.04422},
 primaryClass = {astro-ph.HE},
       adsurl = {https://ui.adsabs.harvard.edu/abs/2017MNRAS.467.4841B}
}

@ARTICLE{JQV22,
       author = {{Quirola-V{\'a}squez}, J. and {Bauer}, F.~E. and {Jonker}, P.~G. and {Brandt}, W.~N. and {Yang}, G. and {Levan}, A.~J. and {Xue}, Y.~Q. and {Eappachen}, D. and {Zheng}, X.~C. and {Luo}, B.},
        title = "{Extragalactic fast X-ray transient candidates discovered by Chandra (2000-2014)}",
      journal = {\aap},
         year = 2022,
        month = jul,
       volume = {663},
          eid = {A168},
        pages = {A168},
          doi = {10.1051/0004-6361/202243047},
archivePrefix = {arXiv},
       eprint = {2201.07773},
 primaryClass = {astro-ph.HE},
       adsurl = {https://ui.adsabs.harvard.edu/abs/2022A&A...663A.168Q}
}

@ARTICLE{JQV23,
       author = {{Quirola-V{\'a}squez}, J. and {Bauer}, F.~E. and {Jonker}, P.~G. and {Brandt}, W.~N. and {Yang}, G. and {Levan}, A.~J. and {Xue}, Y.~Q. and {Eappachen}, D. and {Camacho}, E. and {Ravasio}, M.~E. and {Zheng}, X.~C. and {Luo}, B.},
        title = "{Extragalactic fast X-ray transient candidates discovered by Chandra (2014-2022)}",
      journal = {\aap},
         year = 2023,
        month = jul,
       volume = {675},
          eid = {A44},
        pages = {A44},
          doi = {10.1051/0004-6361/202345912},
archivePrefix = {arXiv},
       eprint = {2304.13795},
 primaryClass = {astro-ph.HE},
       adsurl = {https://ui.adsabs.harvard.edu/abs/2023A&A...675A..44Q}
}

@ARTICLE{lv2014,
       author = {{L{\"u}}, Hou-Jun and {Zhang}, Bing},
        title = "{A Test of the Millisecond Magnetar Central Engine Model of Gamma-Ray Bursts with Swift Data}",
      journal = {\apj},
         year = 2014,
        month = apr,
       volume = {785},
       number = {1},
          eid = {74},
        pages = {74},
          doi = {10.1088/0004-637X/785/1/74},
archivePrefix = {arXiv},
       eprint = {1401.1562},
 primaryClass = {astro-ph.HE},
       adsurl = {https://ui.adsabs.harvard.edu/abs/2014ApJ...785...74L}
}

@ARTICLE{epsilonB,
       author = {{Santana}, Rodolfo and {Barniol Duran}, Rodolfo and {Kumar}, Pawan},
        title = "{Magnetic Fields in Relativistic Collisionless Shocks}",
      journal = {\apj},
         year = 2014,
        month = apr,
       volume = {785},
       number = {1},
          eid = {29},
        pages = {29},
          doi = {10.1088/0004-637X/785/1/29},
archivePrefix = {arXiv},
       eprint = {1309.3277},
 primaryClass = {astro-ph.HE},
       adsurl = {https://ui.adsabs.harvard.edu/abs/2014ApJ...785...29S}
}

@article{Troja_2007,
doi = {10.1086/519450},
url = {https://doi.org/10.1086/519450},
year = {2007},
month = {aug},
publisher = {},
volume = {665},
number = {1},
pages = {599},
author = {Troja, E. and Cusumano, G. and O’Brien, P. T. and Zhang, B. and Sbarufatti, B. and Mangano, V. and Willingale, R. and Chincarini, G. and Osborne, J. P. and Marshall, F. E. and Burrows, D. N. and Campana, S. and Gehrels, N. and Guidorzi, C. and Krimm, H. A. and La Parola, V. and Liang, E. W. and Mineo, T. and Moretti, A. and Page, K. L. and Romano, P. and Tagliaferri, G. and Zhang, B. B. and Page, M. J. and Schady, P.},
title = {Swift Observations of GRB 070110: An Extraordinary X-Ray Afterglow Powered by the Central Engine},
journal = {The Astrophysical Journal}
}

@article{Bucciantini2011,
    author = {Bucciantini, N. and Metzger, B. D. and Thompson, T. A. and Quataert, E.},
    title = {Short gamma-ray bursts with extended emission from magnetar birth: jet formation and collimation},
    journal = {Monthly Notices of the Royal Astronomical Society},
    volume = {419},
    number = {2},
    pages = {1537-1545},
    year = {2011},
    month = {12},
    issn = {0035-8711},
    doi = {10.1111/j.1365-2966.2011.19810.x},
    url = {https://doi.org/10.1111/j.1365-2966.2011.19810.x},
    eprint = {https://academic.oup.com/mnras/article-pdf/419/2/1537/3125386/mnras0419-1537.pdf},
}

@ARTICLE{redback,
       author = {{Sarin}, Nikhil and {H{\"u}bner}, Moritz and {Omand}, Conor M.~B. and {Setzer}, Christian N. and {Schulze}, Steve and {Adhikari}, Naresh and {Sagu{\'e}s-Carracedo}, Ana and {Galaudage}, Shanika and {Wallace}, Wendy F. and {Lamb}, Gavin P. and {Lin}, En-Tzu},
        title = "{REDBACK: a Bayesian inference software package for electromagnetic transients}",
      journal = {\mnras},
         year = 2024,
        month = jun,
       volume = {531},
       number = {1},
        pages = {1203-1227},
          doi = {10.1093/mnras/stae1238},
archivePrefix = {arXiv},
       eprint = {2308.12806},
 primaryClass = {astro-ph.HE},
       adsurl = {https://ui.adsabs.harvard.edu/abs/2024MNRAS.531.1203S}
}

@ARTICLE{JWST_XT2,
       author = {{Quirola-V{\'a}squez}, J. and {Bauer}, F.~E. and {Jonker}, P.~G. and {Levan}, A. and {Brandt}, W.~N. and {Ravasio}, M. and {Eappachen}, D. and {Xue}, Y.~Q. and {Zheng}, X.~C.},
        title = "{New JWST redshifts for the host galaxies of CDF-S XT1 and XT2: Understanding their nature}",
      journal = {\aap},
         year = 2025,
        month = mar,
       volume = {695},
          eid = {A279},
        pages = {A279},
          doi = {10.1051/0004-6361/202451825},
archivePrefix = {arXiv},
       eprint = {2410.10015},
 primaryClass = {astro-ph.HE},
       adsurl = {https://ui.adsabs.harvard.edu/abs/2025A&A...695A.279Q}
}

@ARTICLE{lv2015,
       author = {{L{\"u}}, Hou-Jun and {Zhang}, Bing and {Lei}, Wei-Hua and {Li}, Ye and {Lasky}, Paul D.},
        title = "{The Millisecond Magnetar Central Engine in Short GRBs}",
      journal = {\apj},
         year = 2015,
        month = jun,
       volume = {805},
       number = {2},
          eid = {89},
        pages = {89},
          doi = {10.1088/0004-637X/805/2/89},
archivePrefix = {arXiv},
       eprint = {1501.02589},
 primaryClass = {astro-ph.HE},
       adsurl = {https://ui.adsabs.harvard.edu/abs/2015ApJ...805...89L}
}

@INPROCEEDINGS{IRAF,
       author = {{Tody}, Doug},
        title = "{The IRAF Data Reduction and Analysis System}",
    booktitle = {Instrumentation in astronomy VI},
         year = 1986,
       editor = {{Crawford}, David L.},
       series = {Society of Photo-Optical Instrumentation Engineers (SPIE) Conference Series},
       volume = {627},
        month = jan,
        pages = {733},
          doi = {10.1117/12.968154},
       adsurl = {https://ui.adsabs.harvard.edu/abs/1986SPIE..627..733T}
}

@ARTICLE{solve-field,
       author = {{Lang}, Dustin and {Hogg}, David W. and {Mierle}, Keir and {Blanton}, Michael and {Roweis}, Sam},
        title = "{Astrometry.net: Blind Astrometric Calibration of Arbitrary Astronomical Images}",
      journal = {\aj},
         year = 2010,
        month = may,
       volume = {139},
       number = {5},
        pages = {1782-1800},
          doi = {10.1088/0004-6256/139/5/1782},
archivePrefix = {arXiv},
       eprint = {0910.2233},
 primaryClass = {astro-ph.IM},
       adsurl = {https://ui.adsabs.harvard.edu/abs/2010AJ....139.1782L}
}

@ARTICLE{LCO40_gcn,
       author = {{Selezneva}, A. and {Basurto Merino}, J. and {Berdayes}, P.~G. and {Caballero-Almagro}, A. and {Cer{\'o}n}, A. and {Contreras}, M. and {D{\'\i}az-Segado}, F. and {Ferrer-Lavi{\~n}a}, T. and {Gandolfi}, B. and {Ghiraldo}, V. and {Hern{\'a}ndez Fung}, J. and {Juli{\'a}-Maroto}, L. and {Lekaroz-Urriza}, E. and {Manzano Garc{\'\i}a}, M. and {Mej{\'\i}a-Mart{\'\i}nez}, E. and {Prieto Polo}, J. and {Pulido-Torres}, M. and {Quintana-Ansaldo}, M. and {Schenone-Zanuzzi}, A. and {Tundidor Rodr{\'\i}guez}, T. and {Urquijo-Rodr{\'\i}guez}, E. and {Abdul-Masih}, M. and {P{\'e}rez-Fournon}, I.},
        title = "{EP251023a: ULL-ASTRO-MASTER detection of the optical afterglow with LCO 40-cm telescope at Teide Observatory}",
      journal = {GRB Coordinates Network},
         year = 2025,
        month = oct,
       volume = {42406},
        pages = {1},
       adsurl = {https://ui.adsabs.harvard.edu/abs/2025GCN.42406....1S}
}

@ARTICLE{COLIBRI_gcn,
       author = {{Mandarakas}, Nikos and {Gill}, Ramandeep and {Ducoin}, Jean-Gr{\'e}goire and {Watson}, Alan M. and {Globus}, No{\'e}mie and {Basa}, St{\'e}phane and {Lee}, William H. and {Atteia}, Jean-Luc and {Angulo}, Camila and {Akl}, Dalya and {Antier}, Sarah and {Becerra}, Rosa L. and {Butler}, Nathaniel R. and {Dornic}, Damien and {Fortin}, Francis and {Garc{\'\i}a Garc{\'\i}a}, Leonardo and {Ocelotl L{\'o}pez}, Kin and {L{\'o}pez-C{\'a}mara}, Diego and {Magnani}, Francesco and {Moreno M{\'e}ndez}, Enrique and {Pereyra}, Margarita and {Avo Rakotondrainibe}, Ny and {S{\'a}nchez {\'A}lvarez}, Fredd and {Schneider}, Benjamin and {de Ugarte Postigo}, Antonio},
        title = "{EP251023a: COLIBR{\'I} optical observations}",
      journal = {GRB Coordinates Network},
         year = 2025,
        month = oct,
       volume = {42400},
        pages = {1},
       adsurl = {https://ui.adsabs.harvard.edu/abs/2025GCN.42400....1M}
}

@ARTICLE{UAFO_ABAOgcn,
       author = {{Volnova}, A. and {Pozanenko}, A. and {Kochergin}, A. and {Inasaridze}, R. Ya. and {IKI-GRB-FuN}},
        title = "{EP251023a: UAFO and AbAO optical observations}",
      journal = {GRB Coordinates Network},
         year = 2025,
        month = nov,
       volume = {42716},
        pages = {1},
       adsurl = {https://ui.adsabs.harvard.edu/abs/2025GCN.42716....1V}
}

@ARTICLE{Mondy_gcn,
       author = {{Volnova}, A. and {Pozanenko}, A. and {Pankov}, N. and {Klunko}, E. and {IKI-GRB-FuN}},
        title = "{EP251023a: Mondy optical observations}",
      journal = {GRB Coordinates Network},
         year = 2025,
        month = oct,
       volume = {42468},
        pages = {1},
       adsurl = {https://ui.adsabs.harvard.edu/abs/2025GCN.42468....1V}
}

@ARTICLE{AZT20_gcn,
       author = {{Volnova}, A. and {Pozanenko}, A. and {Krugov}, M. and {IKI-GRB-FuN}},
        title = "{EP251023a: Assy AZT-20 optical observations}",
      journal = {GRB Coordinates Network},
         year = 2025,
        month = nov,
       volume = {42717},
        pages = {1},
       adsurl = {https://ui.adsabs.harvard.edu/abs/2025GCN.42717....1V}
}

@ARTICLE{GIT_gcn,
       author = {{Mohan}, T. and {Eappachen}, D. and {Swain}, V. and {Saikia}, A.~P. and {Bhalerao}, V. and {Anupama}, G.~C. and {Barway}, S. and {Angail}, K. and {GIT Team}},
        title = "{EP251023a: GROWTH-India Telescope optical observations}",
      journal = {GRB Coordinates Network},
         year = 2025,
        month = oct,
       volume = {42411},
        pages = {1},
       adsurl = {https://ui.adsabs.harvard.edu/abs/2025GCN.42411....1M}
}

@ARTICLE{AZT22_gcn,
       author = {{Rajabov}, Y. and {Burkhonov}, O. and {Abidkhanov}, B. and {Ehgamberdiev}, S. and {Tillayev}, Y. and {Boyqobilov}, T. and {Shaymanov}, A. and {UBAI Team}},
        title = "{EP251023a: MAO/AZT-22 optical observations}",
      journal = {GRB Coordinates Network},
         year = 2025,
        month = oct,
       volume = {42422},
        pages = {1},
       adsurl = {https://ui.adsabs.harvard.edu/abs/2025GCN.42422....1R}
}

@ARTICLE{OHP_gcn,
       author = {{Rakotondrainibe}, N.~A. and {Adami}, C. and {Le Floc'h}, E. and {Mistral Grb Collaboration}},
        title = "{EP251023a: OHP/T193 optical observations}",
      journal = {GRB Coordinates Network},
         year = 2025,
        month = oct,
       volume = {42420},
        pages = {1},
       adsurl = {https://ui.adsabs.harvard.edu/abs/2025GCN.42420....1R}
}

@article{GRB070707_zhu,
    author = {Zhu, Yi-Ming and Zhou, Hao and Wang, Yun and Liao, Neng-Hui and Jin, Zhi-Ping and Wei, Da-Ming},
    title = {The afterglow of GRB 070707 and a possible kilonova component},
    journal = {Monthly Notices of the Royal Astronomical Society},
    volume = {521},
    number = {1},
    pages = {269-277},
    year = {2023},
    month = {02},
    issn = {0035-8711},
    doi = {10.1093/mnras/stad541},
    url = {https://doi.org/10.1093/mnras/stad541},
    eprint = {https://academic.oup.com/mnras/article-pdf/521/1/269/49418962/stad541.pdf},
}

@article{Li_2012,
doi = {10.1088/0004-637X/758/1/27},
url = {https://doi.org/10.1088/0004-637X/758/1/27},
year = {2012},
month = {sep},
publisher = {The American Astronomical Society},
volume = {758},
number = {1},
pages = {27},
author = {Li, Liang and Liang, En-Wei and Tang, Qing-Wen and Chen, Jie-Min and Xi, Shao-Qiang and Lü, Hou-Jun and Gao, He and Zhang, Bing and Zhang, Jin and Yi, Shuang-Xi and Lu, Rui-Jing and Lü, Lian-Zhong and Wei, Jian-Yan},
title = {A COMPREHENSIVE STUDY OF GAMMA-RAY BURST OPTICAL EMISSION. I. FLARES AND EARLY SHALLOW-DECAY COMPONENT},
journal = {The Astrophysical Journal}
}

@article{GRB220831A,
    author = {Freeburn, James and O’Connor, Brendan and Cooke, Jeff and Dobie, Dougal and Möller, Anais and Tejos, Nicolas and Zhang, Jielai and Beniamini, Paz and Auchettl, Katie and DeLaunay, James and Dichiara, Simone and Fong, Wen-fai and Goode, Simon and Gordon, Alexa and Kilpatrick, Charles D and Lien, Amy and Mihalenko, Cassidy and Ryan, Geoffrey and Siellez, Karelle and Suhr, Mark and Troja, Eleonora and Van Bemmel, Natasha and Webb, Sara},
    title = {GRB 220831A: a hostless, intermediate gamma-ray burst with an unusual optical afterglow},
    journal = {Monthly Notices of the Royal Astronomical Society},
    volume = {537},
    number = {2},
    pages = {2061-2078},
    year = {2025},
    month = {02},
    issn = {0035-8711},
    doi = {10.1093/mnras/staf147},
    url = {https://doi.org/10.1093/mnras/staf147},
    eprint = {https://academic.oup.com/mnras/article-pdf/537/2/2061/61652452/staf147.pdf},
}

@article{Wang_2015,
doi = {10.1088/0067-0049/219/1/9},
url = {https://doi.org/10.1088/0067-0049/219/1/9},
year = {2015},
month = {jul},
publisher = {The American Astronomical Society},
volume = {219},
number = {1},
pages = {9},
author = {Wang, Xiang-Gao and Zhang, Bing and Liang, En-Wei and Gao, He and Li, Liang and Deng, Can-Min and Qin, Song-Mei and Tang, Qing-Wen and Kann, D. Alexander and Ryde, Felix and Kumar, Pawan},
title = {HOW BAD OR GOOD ARE THE EXTERNAL FORWARD SHOCK AFTERGLOW MODELS OF GAMMA-RAY BURSTS?},
journal = {The Astrophysical Journal Supplement Series}
}

@article{Ronchini_2023,
	author = {{Ronchini}, S. and {Stratta}, G. and {Rossi}, A. and {Kann}, D. A. and {Oganeysan}, G. and {Dall’Osso}, S. and {Branchesi}, M. and {De Cesare}, G.},
	title = {Combined X-ray and optical analysis to probe the origin of the plateau emission in γ-ray burst afterglows},
	DOI= "10.1051/0004-6361/202245348",
	url= "https://doi.org/10.1051/0004-6361/202245348",
	journal = {A\&A},
	year = 2023,
	volume = 675,
	pages = "A117",
}

@article{Li_2026,
doi = {10.3847/1538-4357/ae346b},
url = {https://doi.org/10.3847/1538-4357/ae346b},
year = {2026},
month = {feb},
publisher = {The American Astronomical Society},
volume = {998},
number = {2},
pages = {298},
author = {Li, Xiao-Yan and Liu, Tong and Huang, Bao-Quan and Deng, Chen},
title = {Statistical Analysis of Multiband Plateaus in Gamma-Ray Burst Afterglows},
journal = {The Astrophysical Journal}
}

@article{GRB180618A,
doi = {10.3847/1538-4357/ac972b},
url = {https://doi.org/10.3847/1538-4357/ac972b},
year = {2022},
month = {nov},
publisher = {The American Astronomical Society},
volume = {939},
number = {2},
pages = {106},
author = {Jordana-Mitjans, N. and Mundell, C. G. and Guidorzi, C. and Smith, R. J. and Ramírez-Ruiz, E. and Metzger, B. D. and Kobayashi, S. and Gomboc, A. and Steele, I. A. and Shrestha, M. and Marongiu, M. and Rossi, A. and Rothberg, B.},
title = {A Short Gamma-Ray Burst from a Protomagnetar Remnant},
journal = {The Astrophysical Journal}
}

@article{Rowlinson_2013,
    author = {Rowlinson, A. and O'Brien, P. T. and Metzger, B. D. and Tanvir, N. R. and Levan, A. J.},
    title = {Signatures of magnetar central engines in short GRB light curves},
    journal = {Monthly Notices of the Royal Astronomical Society},
    volume = {430},
    number = {2},
    pages = {1061-1087},
    year = {2013},
    month = {04},
    issn = {0035-8711},
    doi = {10.1093/mnras/sts683},
    url = {https://doi.org/10.1093/mnras/sts683},
    eprint = {https://academic.oup.com/mnras/article-pdf/430/2/1061/9378973/sts683.pdf},
}

@article{Mao_2010,
	author = {{Mao}, Z. and {Yu}, Y. W. and {Dai}, Z. G. and {Pi}, C. M. and {Zheng}, X. P.},
	title = {The termination shock of a magnetar wind: a possible origin of gamma-ray burst X-ray afterglow emission},
	DOI= "10.1051/0004-6361/200913252",
	url= "https://doi.org/10.1051/0004-6361/200913252",
	journal = {A\&A},
	year = 2010,
	volume = 518,
	pages = "A27",
	month = "",
}

@article{Metzger_2014,
    author = {Metzger, Brian D. and Piro, Anthony L.},
    title = {Optical and X-ray emission from stable millisecond magnetars formed from the merger of binary neutron stars},
    journal = {Monthly Notices of the Royal Astronomical Society},
    volume = {439},
    number = {4},
    pages = {3916-3930},
    year = {2014},
    month = {04},
    issn = {0035-8711},
    doi = {10.1093/mnras/stu247},
    url = {https://doi.org/10.1093/mnras/stu247},
    eprint = {https://academic.oup.com/mnras/article-pdf/439/4/3916/3992982/stu247.pdf},
}

@article{Strang_2019,
    author = {Strang, L C and Melatos, A},
    title = {Plerion model of the X-ray plateau in short gamma-ray bursts},
    journal = {Monthly Notices of the Royal Astronomical Society},
    volume = {487},
    number = {4},
    pages = {5010-5018},
    year = {2019},
    month = {08},
    issn = {0035-8711},
    doi = {10.1093/mnras/stz1648},
    url = {https://doi.org/10.1093/mnras/stz1648},
    eprint = {https://academic.oup.com/mnras/article-pdf/487/4/5010/28882188/stz1648.pdf},
}

@article{Dainotti_2022,
doi = {10.3847/1538-4365/ac7c64},
url = {https://doi.org/10.3847/1538-4365/ac7c64},
year = {2022},
month = {jul},
publisher = {The American Astronomical Society},
volume = {261},
number = {2},
pages = {25},
author = {Dainotti, M. G. and Young, S. and Li, L. and Levine, D. and Kalinowski, K. K. and Kann, D. A. and Tran, B. and Zambrano-Tapia, L. and Zambrano-Tapia, A. and Cenko, S. B. and Fuentes, M. and Sánchez-Vázquez, E. G. and Oates, S. R. and Fraija, N. and Becerra, R. L. and Watson, A. M. and Butler, N. R. and González, J. J. and Kutyrev, A. S. and Lee, W. H. and Prochaska, J. X. and Ramirez-Ruiz, E. and Richer, M. G. and Zola, S.},
title = {The Optical Two- and Three-dimensional Fundamental Plane Correlations for Nearly 180 Gamma-Ray Burst Afterglows with Swift/UVOT, RATIR, and the Subaru Telescope},
journal = {The Astrophysical Journal Supplement Series}
}

@ARTICLE{XRTmagnetar,
       author = {{Quirola-V{\'a}squez}, J. and {Bauer}, F.~E. and {Jonker}, P.~G. and {Brandt}, W.~N. and {Eappachen}, D. and {Levan}, A.~J. and {L{\'o}pez}, E. and {Luo}, B. and {Ravasio}, M.~E. and {Sun}, H. and {Xue}, Y.~Q. and {Yang}, G. and {Zheng}, X.~C.},
        title = "{Probing a magnetar origin for the population of extragalactic fast X-ray transients detected by Chandra}",
      journal = {\aap},
         year = 2024,
        month = mar,
       volume = {683},
          eid = {A243},
        pages = {A243},
          doi = {10.1051/0004-6361/202347629},
archivePrefix = {arXiv},
       eprint = {2401.01415},
 primaryClass = {astro-ph.HE},
       adsurl = {https://ui.adsabs.harvard.edu/abs/2024A&A...683A.243Q}
}

@article{Campana2010,
    author = {Campana, S. and Thöne, C. C. and de Ugarte Postigo, A. and Tagliaferri, G. and Moretti, A. and Covino, S.},
    title = {The X-ray absorbing column densities of Swift gamma-ray bursts},
    journal = {Monthly Notices of the Royal Astronomical Society},
    volume = {402},
    number = {4},
    pages = {2429-2435},
    year = {2010},
    month = {03},
    issn = {0035-8711},
    doi = {10.1111/j.1365-2966.2009.16006.x},
    url = {https://doi.org/10.1111/j.1365-2966.2009.16006.x},
    eprint = {https://academic.oup.com/mnras/article-pdf/402/4/2429/4890893/mnras0402-2429.pdf},
}

\begin{appendix}
\nolinenumbers



\section{X-ray data reduction and analysis}
The WXT data were processed using the WXT Data Analysis Software (WXTDAS) with the latest calibration database. The photons of the source and the background were extracted from a circle with a radius of $9'$ and an annulus with radii of $18'$ and $36'$, respectively. As the WXT average net count rate is $\sim 0.5 \, \rm cts\,s^{-1}$ in the total prompt emission phase, we grouped the WXT data with 3 minimum counts per bin to perform the spectral analysis. We fit the WXT data with the absorbed power-law model (\texttt{TbAbs$\times$zTbAbs$\times$PowerLaw}). The first and second components are responsible for Galactic absorption and intrinsic absorption using the T$\mathrm{\ddot{u}}$ebingen-Boulder interstellar medium absorption model.
For the Galactic hydrogen column density, we adopted $N_{\rm{H}} = 4.2 \times 10^{20}\,\rm{cm}^{-2}$ throughout the Letter, as calculated by the UK Swift Science Data Centre\footnote{\url{https://www.swift.ac.uk/analysis/nhtot/}}.
The third component is a simple photon power law, with $N(E)=K(\frac{E}{1\,\rm{keV}})^{-\Gamma}$, where $K$ is the normalization of the spectrum and $\Gamma$ is the dimensionless photon index of the power law. The results are presented in Table \ref{Xray_result}

The FXT cleaned event ﬁles and response files were generated by using the Follow-up X-ray Telescope Data Analysis Software (\texttt{FXTDAS v1.10}\footnote{\url{http://epfxt.ihep.ac.cn/analysis}}). With the $90\%$ of the Point Spread Function (PSF) is enclosed by a $\sim 1'$ radius circle at 1.5 keV, the photons of source and background were extracted from a circle with a radius of $1'$ and an annulus with radii of $2'$ and $3'$, respectively. The FXT data at various times were also grouped with different counts to enhance the signal-to-noise ratio.

\section{Optical data reduction and analysis}\label{sec:optical_details}
\begin{figure}[htbp!]
\centering
\includegraphics[width=0.4\textwidth, keepaspectratio]{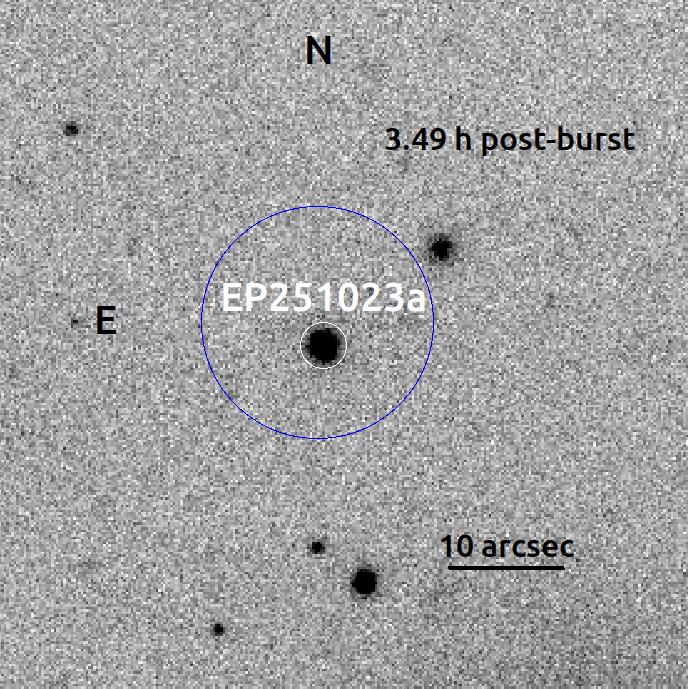}
\caption{The Sloan $r'$-band position of EP251023a with the FOV $1^{'} \times 1^{'}$ obtained by NOT/ALFOSC about 3.49 h after the discovery of EP251023a. The location of the burst is circled in white and the 10$\arcsec$ error circle of EP/FXT is in blue.
\label{locimg}}
\end{figure}

The celestial location of the burst is shown in Fig.~\ref{locimg}. The follow-up observations have been performed by a considerable number of ground-based telescopes. In addition to the publicly available data from the GCN circulars, we include new observations obtained with the following telescopes: the 2.56~m Nordic Optical Telescope (NOT; located at the Roque de los Muchachos Observatory, La Palma, Spain); the 1~m telescope at the Las Cumbres Observatory (LCO; through the Global Supernova Project; Global); the 0.6~m robotic telescope of Burst Observer and Optical Transient Exploring System (BOOTES-7; located at San Pedro de Atacama, Chile); the 1~m JinShan 100A (C) telescopes (ALT/100A and ALT/100C; located at Altay Observatory, Xinjiang, China). 
After standard data reduction with \texttt{IRAF} \citep{IRAF}, and astrometric calibration by Astrometry.net \citep{solve-field}, the optical photometry was calibrated with the nearby Pan-STARRS1 DR2 reference stars. The photometry in Johnson–Cousin filters were calibrated with the converted magnitude from the Sloan system\footnote{\url{https://www.sdss4.org/dr12/algorithms/sdssUBVRITransform/}} for the nearby reference stars. The photometric results are presented in Table~\ref{tab:optical_result}.

To construct SEDs at multiple epochs, we aligned the optical data with the X-ray observations by extrapolating the nearest optical data points to the two X-ray epochs using the measured temporal decay slopes. Among these, extensive multi-wavelength coverage is available at the epoch marked by the third blue vertical line ($\sim$ 96.3 ks) in Fig.~\ref{totallc}. We therefore fit the SED at 96.3 ks with an absorbed power-law using the \texttt{zDust$\times$TbAbs$\times$zTbAbs$\times$PowerLaw} model in \texttt{Xspec v12.14.0h}, where \texttt{zDust} represents extinction by dust grains in the host galaxy of the burst, and \texttt{TbAbs} and \texttt{zTbAbs} are, respectively, hydrogen photoelectric absorption in the Milky Way Galaxy and the host galaxy. The redshift and the Galactic hydrogen column density are fixed to 2.232 and $N_{\rm{H}} = 4.2 \times 10^{20}\,\rm{cm}^{-2}$, respectively. For all three extinction laws (Milky Way, Large Magellanic Cloud, or Small Magellanic Cloud), $E(B - V)$ of the host galaxy cannot be accurately constrained and tends toward zero under the optimal statistical conditions. Thus, host-galaxy extinction is negligible.

\begin{figure*}[htbp!]
\includegraphics[width=1.0\textwidth, keepaspectratio]{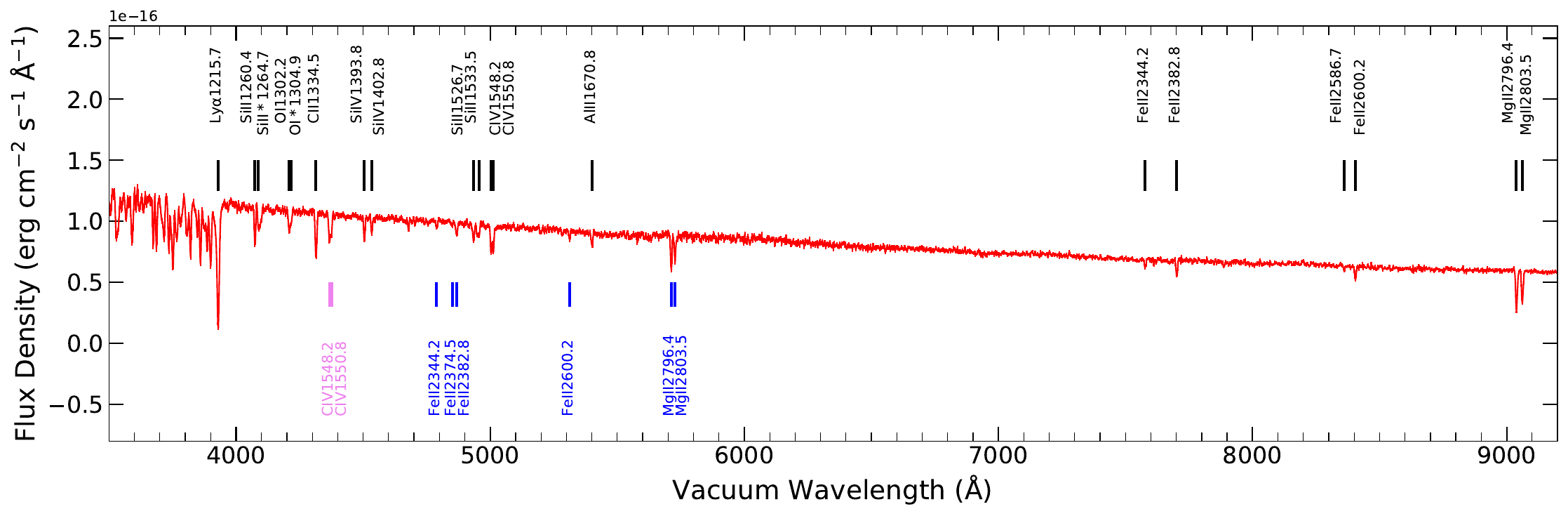}
\caption{The spectrum obtained with Keck/LRIS. The identified metal absorption lines are indicated with vertical dashes. Black dashes represent the absorption lines from the host galaxy at $z=2.232$, blue dashes indicate the foreground system at $z=1.043$, and the violet dashes mark the possible C IV lines at $z=1.882$.
\label{optspec}}
\end{figure*}

\begin{figure}[htbp!]
\includegraphics[width=0.45\textwidth, keepaspectratio]{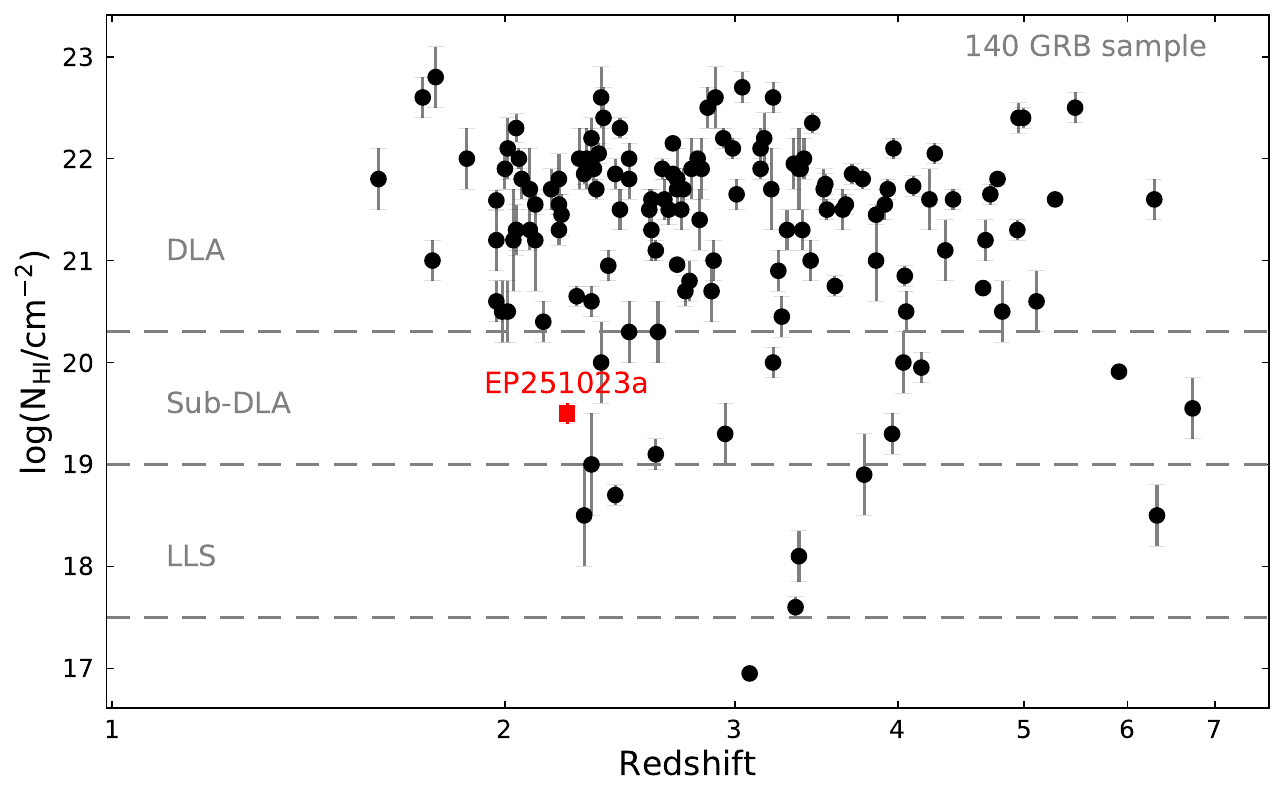}
\caption{The values of neutral hydrogen column density in the host plotted against redshift for the sample of 140 GRBs \citep{2019MNRAS.483.5380T}. The vast majority (123 out of 140) are DLAs, with the remainder classified as as sub-DLAs, Lyman Limit Systems (LLS) and below. The neutral hydrogen column density of EP251023a are illustrated as red square.
\label{NH_fig}}
\end{figure}

For Keck/LRIS, the spectrum was acquired with the slit oriented near the parallactic angle to minimize slit losses caused by atmospheric dispersion. The LRIS observations utilized the $1\arcsec$-wide slit, 600/4000 grism, and 400/8500 grating, which produced a spectral coverage of 3140--10,270\,\r{A}. Data reduction followed standard techniques for CCD processing and spectrum extraction. Low-order polynomial fits to comparison-lamp spectra were used to calibrate the wavelength scale, and small adjustments derived from night-sky lines in the target frames were applied. The spectrum was flux calibrated using observations of appropriate spectrophotometric standard stars observed on the same night, at similar airmasses, and with an identical instrument configuration; these standard-star spectra were also used to remove telluric absorption. 
The spectrum of EP251023a was acquired with exposure times of 950 s and $2 \times 450$ s for the blue and red sides, respectively. Numerous strong absorption lines --including Lyman--alpha, Si II, Si II*, O I, O I*, C II, Si IV, C IV, Al II, Fe II, and Mg II -- yield a consistent redshift of $z = 2.232 \pm 0.001$, and some Fe II and Mg II absorption lines identified a foreground system at $z=1.043 \pm 0.001$. The absorption doublet near $3700 \AA$ may correspond to the C IV doublet at $z=1.822$, but remains unconfirmed due to the absence of additional absorption lines.
The ﬁt to the strong Ly$\alpha$ absorption line yields a column density of log~($N_{\rm HI}/cm^{-2}$) = $19.5 \pm 0.1$. According to \cite{2019MNRAS.483.5380T}, neutral hydrogen column densities measured in GRB host galaxies are typically high, mostly classified as damped Ly$\alpha$ absorbers (DLAs; log~($N_{\rm HI}/cm^{-2}$)$\,>\,20.3$). This suggests that type II GRB progenitors reside in dense molecular clouds or the gas-rich central regions of star-forming galaxies. The neutral hydrogen column density of EP251023a is classified as sub-DLA ($19\,<\,$log~($N_{\rm HI}/cm^{-2}$)$\,<\,20.3$), which is relatively low compared to those typically found in GRB host galaxies. In Fig.~\ref{NH_fig}, we compared the result with 140 other GRB samples from \cite{2019MNRAS.483.5380T}. The intrinsic hydrogen column density inferred from the X-ray data (a few $\times10^{22}\rm cm^{-2}$) significantly exceeds that derived from the optical spectrum, a common feature in Swift GRBs \citep{Campana2010} that may result from the ionization of local hydrogen by the burst.

\section{The physical origin of the afterglow emission}\label{sec:origin appendix}
If the optical afterglow arises from the forward shock (as predicted by the standard fireball model), the observed steepening may signify a jet break. However, jet break would require $\alpha_2=p$ \citep{1999ApJ...525..737R}, where $p$ is the electron energy distribution index. Such a steep decay could be marginally consistent with the post-break phase only by adopting a very soft electron energy distribution ($p>3$); this is unusual in observations \citep{Wang_2015}. Moreover, the magnitude of the steepening from the pre- to the post-break decay ($\Delta \alpha=\alpha_2-\alpha_1\approx3.7$) would be too pronounced for a jet-break interpretation.

A steep decline in the ambient density can also suppress the afterglow emission in the forward-shock model. According to \cite{density_jump}, for a density contrast of $\sim$10, the maximum changes of $\Delta \alpha \approx 0.4$, increasing to the maximum changes of $\Delta \alpha \approx 1$ at very large density contrasts (e.g., $\sim$1000). Hence, this scenario is not suitable to explain such an extremely steep decay in EP251023a.

An external-shock afterglow is generally expected in GRB-like transients; however, no clear evidence for such emission is detected in EP251023a. We therefore discuss the lack of observed external forward shock emission.
According to the conservative $E_{\gamma,\rm{iso}}$ upper limit of $5.7 \times 10^{52}$ erg, we take the isotropic kinetic energy $E_{\rm k,iso}=1\times10^{53} \rm erg$. The typical values of the jet opening angle $\theta_{j}=0.1$ rad and $p=2.3$ are adopted. For the microphysical afterglow parameters, we assume plausible values of log~$\epsilon_e=-1$ and log~$\epsilon_B=-3.5$ guided by results from previous GRB studies \citep{epsilonB}. Assuming that any underlying forward-shock emission must remain subdominant compared to the observed optical and X-ray emission, we constrain the surrounding medium density by requiring the forward-shock flux to remain below $\sim10\%$ of the observed flux. Under these assumptions, we obtain an upper limit of $n < 5\times10^{-4}\,\rm cm^{-3}$. Such a low circumburst density is also broadly consistent with the relatively low host-galaxy HI column density inferred from the Keck spectrum, although the two quantities probe different physical scales and are not necessarily directly correlated. Notably, the $E_{\gamma,\rm{iso}}$ limit provided by KW is highly conservative. EP251023a is actually expected to have a lower $E_{\rm k,iso}$, which would relax the constraints on the density $n$. This, in turn, makes the non-detection of the external forward shock emission much more plausible.

The characteristic spin-down luminosity $L_0$ and the characteristic spin-down timescale $\tau$ are related to the magnetar initial parameters as

\begin{equation}
L_0=1.0\times10^{49}\rm{ erg\,s^{-1}} (B_{p,15}^{2}P_{0,-3}^{-4}R_6^6)
\end{equation}

\begin{equation}
\tau=2.05\times10^{3}\rm{s} (I_{45}B_{p,15}^{-2}P_{0,-3}^{2}R_6^{-6})
\end{equation}
where $I_{45}$ is the moment of inertia in units of $10^{45} \rm g\,cm^{2}$, $B_{p,15}$ is the magnetic field strength in units of $10^{15}$ G, $P_{0,-3}$ is the initial period in milliseconds, and $R_6$ is stellar radius in units of $10^6$ cm. Here we adopt $R_6=1$ and $I_{45}=1$. The convention Q=$10^xQ_x$ is adopted in cgs units for all other parameters throughout this letter. The spin-down luminosity $L_0$ is related to the isotropic plateau luminosity ($L_{\rm em}$) as 
\begin{equation}
\eta L_0=L_{\rm em}f_b
\end{equation}
where $\eta=\frac{L_{\rm em}}{L_{\rm em}+L_{\rm K}}$ is the radiation efficiency, and $f_b=1-\cos{\theta_j}$ is the beaming factor. As discussed in \cite{lv2014}, the $L_{\rm em}$ can be measured from the flux of the plateau and the $L_{\rm K}$ can be derived from the afterglow. The afterglow of EP251023a is not detected throughout the observation period. Thus, we adopt $\eta=1$, since the afterglow contribution is negligible.
The $L_{\rm em}$ of the plateau can be described as $L_{\rm em,0}$, where:
\begin{equation}\label{eq5}
L_{\rm em}(t)=L_{\rm em,0} \frac{1}{(1+\frac{t}{\tau})^2} \simeq 
\begin{cases} 
L_{\rm em,0}, & t \ll \tau, \\ 
L_{\rm em,0} \left(\dfrac{t}{\tau}\right)^{-2}, & t \gg \tau. 
\end{cases}
\end{equation}
Equation \ref{eq5} describes the luminosity evolution in the electromagnetic spin-down–dominated regime, predicting a post-break decay slope of $\alpha_2 \approx 2$. The spin-down timescale $\tau$ can thus be inferred from the observed evolution. However, a post-break slope steeper than 3 requires an abrupt shutdown of the central engine, likely due to the collapse of a supra-massive magnetar into a black hole before substantial spin-down occurs (first seen in \citealt{Troja_2007}). In this scenario, the observed break time corresponds to the onset of collapse, implying that $\tau$ exceeds $t_{b}/(1+z)$ and thus provides a lower limit.

\section{Comparison with similar transients}\label{sec:comparison appendix}
\begin{figure}[htbp!]
\includegraphics[width=0.45\textwidth, keepaspectratio]{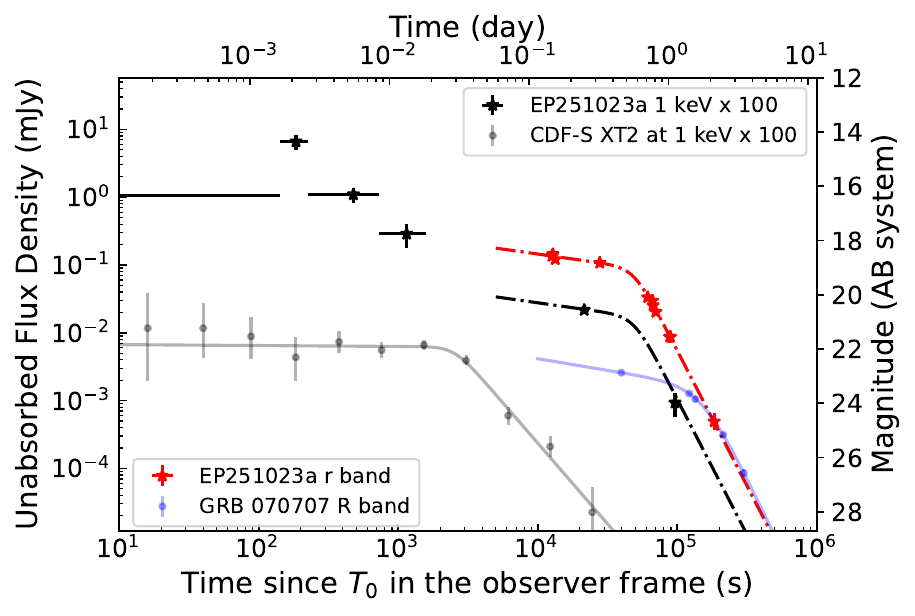}
\caption{The figure illustrates the optical data of EP251023a (red stars) and GRB\,070707 (blue circles), along with the X-ray data of EP251023a (black stars) and CDF-S XT2 (grey circles). The corresponding SBPL models are shown as dash-dotted lines for EP251023a and solid lines for the other two transients.
\label{3Compare}}
\end{figure}

For a comprehensive comparison, we analyze EP251023a in the context of other transients associated with millisecond magnetars. We begin with CDF-S~XT2, an eFXT detected in \textit{Chandra} archival data. It was the first eFXT interpreted as magnetar spin-down emission, with its X-ray light curve exhibiting a plateau ($\alpha_1=0.14 \pm 0.03$) followed by a post-break decay of $\alpha_2=2.16^{+0.29}_{-0.26}$. 
This model fits CDF-S XT2 exceptionally well, notably with a post-plateau decay slope of 2.16 that matches the magnetic dipole radiation profile shown in Equation~\ref{eq5}. However, alternative interpretations for the observed data remain viable, such as the afterglow of a slightly off-axis long GRB or a low-luminosity GRB \citep{JWST_XT2}. Recently, \textit{JWST} imaging and spectroscopy observations \citep{JWST_XT2} have established the redshift of CDF-S XT2 as $3.4598\pm0.0022$, whereas it was previously misidentified as 0.738. We derive $L_{\rm em}=(1.21\pm0.27)\times 10^{47} \,\rm erg\,s^{-1}$ and $\tau = 465\pm81$ s from the observed data for CDF-S XT2. 
As CDF-S XT2 was identified from archival data, multi-band follow-up observations are not available. Given that its X-ray emission is clearly dominated by the magnetar component, we adopt $\eta=1$ for CDF-S XT2 to ensure a consistent comparison with EP251023a, resulting in $P_0=19.09\pm2.70$~ms and $B_p=(4.01\pm1.22)\times10^{16}$~G with the isotropic assumption.
CDF-S XT2 shows a spectral softening trend after the break, whereas EP251023a shows no significant spectral evolution across the broadband SED. While the plateau luminosities of both sources are comparable, the plateau duration of CDF-S XT2 is about two orders of magnitude shorter than that of EP251023a. The X-ray data for these two transients are presented in Fig.~\ref{3Compare}.

Following the identification of the magnetar model for CDF-XT2, \cite{XRTmagnetar} investigated additional eFXTs that might be associated with magnetars. They found that five additional eFXTs support the free-zone scenario, and three others support the trapped-zone scenario. Fig.~\ref{Magnetar compare} displays the two eFXTs with known redshifts. Following the fitting results in \cite{JQV23}, and applying the same assumptions used for CDF-S XT2, XRT\,170901 yields $P_0=57.07\pm5.48$~ms and $B_p=(8.81\pm1.79)\times10^{16}$~G. Regarding XRT\,210423, two potential host galaxy redshifts ($z=1.04$ and $1.5105$) have been proposed. Given its post-break decay slope $ \alpha_2=3.8\pm1.2 $, we derived the corresponding parameter limits for these two redshifts as $P_0<70.73$~ms, $B_p<6.90\times10^{16}$~G and $P_0<49.41$~ms, $B_p<5.34\times10^{16}$~G for two plausible redshifts, respectively. These two eFXTs exhibit isotropic plateau luminosities ($L_{\rm em}$) of a few $\times 10^{45} \,\rm erg\,s^{-1}$, which is more than one order of magnitude lower than those of CDF-S XT2 and EP251023a. The spin-down timescales are $\tau \sim 861$ s for XRT\,170901, and $\tau > 2157$ and $\tau > 1753$ s for the two redshifts of XRT\,210423, respectively. Unlike CDF-S XT2, their X-ray spectra show no evidence of evolution across the break, a feature consistent with EP251023a.

As discussed above, GRB\,070707 remains with optical data similar to EP251023a. We fit the host-subtracted data of GRB\,070707 \citep{GRB070707_zhu} using a SBPL function with $\omega=1$, following the approach of \cite{GRB070707}. We derived $\alpha_1=0.34 \pm 0.05$, $\alpha_2=4.31 \pm 0.15$ and $t_b=153552 \pm 567$ s for GRB\,070707. Due to the uncertain redshift, $L_{\rm em}$ and $\tau$ of GRB\,070707 are evaluated for plausible redshifts between $z=0.24$ and $z=3.6$. The corresponding initial spin period $P_0$ and surface magnetic field $B_p$ are presented in Fig.~\ref{Magnetar compare}. We adopted the isotropic wind scenario, given that the event is a short GRB detected by INTEGRAL. 
Both sources are characterized by a long-duration plateau, a feature robust against redshift uncertainties. In contrast, their plateau luminosities cannot be directly compared, as the unconstrained redshift of GRB\,070707 introduces a large uncertainty in its derived luminosity. X-ray observations of GRB\,070707 are even sparser than those of EP251023a. The only constraint we can derive is that the optical data is inconsistent with the flux density extrapolated from the pre-break X-ray spectrum. The optical data for these two transients are presented in Fig.~\ref{3Compare}.

\section{Acknowledgements}
This work is based on the data obtained with Einstein Probe, a space mission supported by the Strategic Priority Program on Space Science of Chinese Academy of Sciences, in collaboration with the European Space Agency, the Max-Planck-Institute for extraterrestrial Physics (Germany), and the Centre National d'Études Spatiales (France).
Based on observations made with the Nordic Optical Telescope, owned in collaboration by the University of Turku and Aarhus University, and operated jointly by Aarhus University, the University of Turku and the University of Oslo, representing Denmark, Finland and Norway, the University of Iceland and Stockholm University at the Observatorio del Roque de los Muchachos, La Palma, Spain, of the Instituto de Astrofisica de Canarias. The NOT data were obtained under program ID P72-811.
Some of the data presented herein were obtained at Keck Observatory, which is a private 501(c)3 non-profit organization operated as a scientific partnership among the California Institute of Technology, the University of California, and the National Aeronautics and Space Administration. The Observatory was made possible by the generous financial support of the W. M. Keck Foundation. 
The authors wish to recognize and acknowledge the very significant cultural role and reverence that the summit of Maunakea has always had within the Native Hawaiian community. We are most fortunate to have the opportunity to conduct observations from this mountain. 
This work makes use of observations from the Las Cumbres Observatory network.
The LCO team is supported by NSF grants AST-2308113 and AST-1911151.
"D.S., D.F., A.R., A.L.L., A.T., and M.U. was
supported by the basic funding program of the Ioffe Institute no. FFUG-2024-0002.
AJCT acknowledges support from the Spanish Ministry project PID2023-151905OB-I00 and Junta de Andaluc\'ia grant P20\_010168

\onecolumn
\section{X-ray and optical data tables}\label{sec:tables}
\begin{table*}[!htbp]
\caption {Spectral Fitting Results and Corresponding Fitting Statistics for EP \newline}
\label{Xray_result}
\centering
\begin{tabular}{ccccc}
\hline\hline             
Time Intervals & Instruments & Intrinsic Absorption\tablefootmark{a} & Photon Index\tablefootmark{b} & CSTAT/(d.o.f.)\\ 
(second) &  & (cm$^{-2}$) & ($\Gamma$)  &  \\
\hline
0 - 380 & WXT & $(2.13 \pm 1.35) \times 10^{22}$ & $1.77 \pm 0.32$ & 43.08/50\\
0 - 57 & WXT & $--$ & $0.88 \pm 0.51$ &8.70/11\\
57 - 197 & WXT & $--$ & $1.51 \pm 0.26$ &  32.96/40\\
197 - 380 & WXT & $--$ & $2.45 \pm 0.35$ &  27.18/32 \\
380 - 1571 & WXT & $<2.93\times10^{22}$ & $1.25 \pm 0.43$ &  18.75/24 \\
\hline
    2.14 $\times 10^4$ - 2.36 $\times 10^4$ & FXT & $(1.67 \pm 1.15) \times 10^{22}$ & $2.37 \pm 0.31$ & 138.70/140 \\
    9.62 $\times 10^4$ - 9.91 $\times 10^4$ & FXT & $<6.72\times10^{22}$ & $1.28 \pm 0.73$ & 16.17/10 \\
\hline
\end{tabular}
\tablefoot{All error bars represent $1\sigma$ uncertainties. \\
\tablefoottext{a}{Dashes ($--$) in the Intrinsic Absorption indicate that we use the same value as the result above.}\\
\tablefoottext{b}{An absorbed power-law model (\texttt{TbAbs$\times$zTbAbs$\times$PowerLaw}) is used to fit the X-ray data, and the Galactic hydrogen column density is fixed with $N_{\rm{H}} = 4.2 \times 10^{20}\,{\rm cm}^{-2}$. }
}
\end{table*}

\begin{longtable}{ccccc}

\caption{The Photometric Results of Our Observations
Combined with Collected GCN Results}\\
\hline\hline
\label{tab:optical_result} 

$\Delta T$ (day) & Band & Magnitude (AB) & 
Telescope & Reference\\
\hline
\endfirsthead
\caption{continued.}\\
\hline\hline
$\Delta T$ (day) & Band & Magnitude (AB) & 
Telescope & Reference\\
\hline
\endhead
\hline
\endfoot
0.1220&  $L$&  18.57$\pm$0.06&  GOTO&  \citealt{GOTOgcn}  \\
0.1295&  $L$&  18.64$\pm$0.04&  GOTO&  \citealt{GOTOgcn}  \\
0.1466&  $r$&  18.57$\pm$0.01&  NOT&  This work  \\
0.1473&  $r$&  18.50$\pm$0.05&  LCO-100&  This work  \\
0.1546&  $r$&  18.69$\pm$0.10&  LCO-40&  \citealt{LCO40_gcn}  \\
0.1562&  $z$&  18.36$\pm$0.01&  NOT&  This work  \\
0.1720&  $G$&  18.91$\pm$0.07&  BOOTES--7&  This work  \\
0.1978&  $G$&  18.90$\pm$0.06&  BOOTES--7&  This work  \\
0.2256&  $G$&  18.94$\pm$0.04&  BOOTES--7&  This work  \\
0.2514&  $G$&  19.00$\pm$0.04&  BOOTES--7&  This work  \\
0.3220&  $r$&  18.82$\pm$0.01&  COLIBRI&  \citealt{COLIBRI_gcn}  \\
0.3220&  $z$&  18.52$\pm$0.01&  COLIBRI&  \citealt{COLIBRI_gcn}  \\
0.6801&  $R$&  20.15$\pm$0.22&  UAFO/RC-500&  \citealt{UAFO_ABAOgcn}  \\
0.7109&  $r$&  20.09$\pm$0.08&  ALT100A&  This work  \\
0.7543&  $R$&  20.82$\pm$0.12&  Mondy&  \citealt{Mondy_gcn}  \\
0.7686&  $r$&  20.22$\pm$0.06&  AZT-20&  \citealt{AZT20_gcn}  \\
0.7762&  $r$&  20.39$\pm$0.09&  GIT&  \citealt{GIT_gcn}  \\
0.7946&  $g$&  20.68$\pm$0.06&  ALT100C&  This work  \\
0.7978&  $i$&  20.24$\pm$0.11&  GIT&  \citealt{GIT_gcn}  \\
0.8125&  $r$&  20.63$\pm$0.07&  ALT100C&  This work  \\
0.8191&  $g$&  20.88$\pm$0.10&  GIT&  \citealt{GIT_gcn}  \\
0.8303&  $i$&  20.40$\pm$0.07&  ALT100C&  This work  \\
0.8499&  $z$&  20.40$\pm$0.20&  ALT100C&  This work  \\
0.8605&  $R$&  20.74$\pm$0.03&  AZT-22&  \citealt{AZT22_gcn}  \\
1.0207&  $r$&  21.55$\pm$0.22&  LCO-100&  This work  \\
1.0284&  $g$&  21.81$\pm$0.24&  LCO-100&  This work  \\
1.0284&  $i$&  21.29$\pm$0.23&  LCO-100&  This work  \\
1.0337&  $r$&  21.55$\pm$0.10&  OHP/T193&  \citealt{OHP_gcn}  \\
1.7771&  $g$&  $>$22.40&  ALT100C&  This work  \\
1.8090&  $r$&  $>$22.30&  ALT100C&  This work  \\
1.8463&  $i$&  $>$22.10&  ALT100C&  This work  \\
1.8826&  $R$&  $>$20.68&  AbAO/AS-32&  \citealt{UAFO_ABAOgcn}  \\
2.1301&  $r$&  24.67$\pm$0.31&  NOT&  This work  \\
2.7592&  $R$&  $>$22.88&  Mondy&  \citealt{Mondy_gcn}  \\
2.7873&  $r$&  $>$22.80&  ALT100C&  This work  \\
4.1413&  $r$&  $>$24.50&  NOT&  This work  \\
8.1218&  $r$&  $>$25.20&  NOT&  This work  \\
\end{longtable}
\tablefoot{$\Delta T$ is the exposure median time after the $T_0$. Magnitudes in the AB system are not corrected for Galactic extinction, which is $E(B - V) = 0.037$ \citep{SF2011}.}
\twocolumn

\end{appendix}
\end{document}